
%
\def\unredoffs{}
\tolerance=1000\hfuzz=2pt
\catcode`\@=11 
\ifx\hyperdef\UNd@FiNeD\def\hyperdef#1#2#3#4{#4}\def\hyperref#1#2#3#4{#4}\def\href#1#2{#2}\fi
\magnification=1200\unredoffs\baselineskip=16pt plus 2pt minus 1pt
\def\Date#1{\vfill\leftline{#1}\tenpoint\supereject%
\footline={\hss\tenrm\hyperdef\hypernoname{page}\folio\folio\hss}}%

{\count255=\time\divide\count255 by 60 \xdef\hourmin{\number\count255}
 \multiply\count255 by-60\advance\count255 by\time
 \xdef\hourmin{\hourmin:\ifnum\count255<10 0\fi\the\count255}
}
\def\date{\number\day.\number\month.\number\year\ at \hourmin}


\def\nolabels{\def\wrlabeL##1{}\def\eqlabeL##1{}\def\reflabeL##1{}}
\def\writelabels{\def\wrlabeL##1{\leavevmode\vadjust{\rlap{\smash%
{\line{{\escapechar=` \hfill\rlap{\sevenrm\hskip.03in\string##1}}}}}}}%
\def\eqlabeL##1{{\escapechar-1\rlap{\sevenrm\hskip.05in\string##1}}}%
\def\reflabeL##1{\noexpand\llap{\noexpand\sevenrm\string\string\string##1}}}
\nolabels

\global\newcount\secno \global\secno=0
\global\newcount\meqno \global\meqno=1
\def\s@csym{}

\def\newsec#1\par{\global\advance\secno by1%
{\toks0{#1}\message{(\the\secno. \the\toks0)}}%
\global\subsecno=0\eqnres@t\let\s@csym\secsym\xdef\secn@m{\the\secno}\noindent
{\bf\hyperdef\hypernoname{section}{\the\secno}{\the\secno.} #1}%
\writetoca{{\string\hyperref{}{section}{\the\secno}{\bf \the\secno\quad}} {\bf #1}}\par%
\nobreak\medskip\nobreak\noindent\ignorespaces}
\def\eqnres@t{\xdef\secsym{\the\secno.}\global\meqno=1\bigbreak\bigskip}
\def\sequentialequations{\def\eqnres@t{\bigbreak}}\xdef\secsym{}

\global\newcount\subsecno \global\subsecno=0
\def\subsec#1\par{\global\advance\subsecno by1%
{\toks0{#1}\message{(\s@csym\the\subsecno. \the\toks0)}}%
\global\subsubsecno=0%
\ifnum\lastpenalty>9000\else\bigbreak\fi
\noindent{\it\hyperdef\hypernoname{subsection}{\secn@m.\the\subsecno}%
{\secn@m.\the\subsecno.} #1}\writetoca{\string\hskip1.45cm
{\string\hyperref{}{subsection}{\secn@m.\the\subsecno}{\secn@m.\the\subsecno.}}
{#1}}\par\nobreak\medskip\nobreak\noindent\ignorespaces}

\def\appendix#1#2{\global\meqno=1\global\subsecno=0\xdef\secsym{\hbox{#1.}}%
\bigbreak\bigskip\noindent{\bf Appendix \hyperdef\hypernoname{appendix}{#1}%
{#1.} #2}{\toks0{(#1. #2)}\message{\the\toks0}}%
\xdef\s@csym{#1.}\xdef\secn@m{#1}%
\writetoca{{\string\hyperref{}{appendix}{#1}{\bf {#1}\quad}} {\bf #2}}%
\par\nobreak\medskip\nobreak}

%
\def\checkm@de#1#2{\ifmmode{\def\f@rst##1{##1}\hyperdef\hypernoname{equation}%
{#1}{#2}}\else\hyperref{}{equation}{#1}{#2}\fi}
\def\eqnn#1{\DefWarn#1\xdef #1{(\noexpand\relax\noexpand\checkm@de%
{\s@csym\the\meqno}{\secsym\the\meqno})}%
\wrlabeL#1\writedef{#1\leftbracket#1}\global\advance\meqno by1}
\def\f@rst#1{\c@t#1a\em@ark}\def\c@t#1#2\em@ark{#1}
\def\eqna#1{\DefWarn#1\wrlabeL{#1$\{\}$}%
\xdef #1##1{(\noexpand\relax\noexpand\checkm@de%
{\s@csym\the\meqno\noexpand\f@rst{##1}1}{\hbox{$\secsym\the\meqno##1$}})}
\writedef{#1\numbersign1\leftbracket#1{\numbersign1}}\global\advance\meqno by1}
\def\eqn#1#2{\DefWarn#1%
\xdef #1{(\noexpand\hyperref{}{equation}{\s@csym\the\meqno}%
{\secsym\the\meqno})}$$#2\eqno(\hyperdef\hypernoname{equation}%
{\s@csym\the\meqno}{\secsym\the\meqno})\eqlabeL#1$$%
\writedef{#1\leftbracket#1}\global\advance\meqno by1}
\def\xeqn{\expandafter\xe@n}\def\xe@n(#1){#1}
\def\xeqna#1{\expandafter\xe@n#1}
\def\eqns#1{(\e@ns #1{\hbox{}})}
\def\e@ns#1{\ifx\UNd@FiNeD#1\message{eqnlabel \string#1 is undefined.}%
\xdef#1{(?.?)}\fi{\let\hyperref=\relax\xdef\next{#1}}%
\ifx\next\em@rk\def\next{}\else%
\ifx\next#1\xeqn#1\else\def\n@xt{#1}\ifx\n@xt\next#1\else\xeqna#1\fi
\fi\let\next=\e@ns\fi\next}

\def\DefWarn#1{\ifx\UNd@FiNeD#1\else
\immediate\write16{*** WARNING: the label \string#1 is already defined ***}\fi}
%
\newskip\footskip\footskip14pt plus 1pt minus 1pt 
\def\footnotefont{\ninepoint}\def\f@t#1{\footnotefont #1\@foot}
\def\f@@t{\baselineskip\footskip\bgroup\footnotefont\aftergroup\@foot\let\next}
\setbox\strutbox=\hbox{\vrule height9.5pt depth4.5pt width0pt}
\global\newcount\ftno \global\ftno=0
\def\foot{\global\advance\ftno by1\def\foot@rg{\hyperref{}{footnote}%
{\the\ftno}{\the\ftno}\xdef\foot@rg{\noexpand\hyperdef\noexpand\hypernoname%
{footnote}{\the\ftno}{\the\ftno}}}\footnote{$^{\foot@rg}$}}
%
%
%
\global\newcount\refno \global\refno=1
\newwrite\rfile
\def\ref{[\hyperref{}{reference}{\the\refno}{\the\refno}]\nref}
\def\nref#1{\DefWarn#1%
\xdef#1{[\noexpand\hyperref{}{reference}{\the\refno}{\the\refno}]}%
\writedef{#1\leftbracket#1}%
\ifnum\refno=1\immediate\openout\rfile=\jobname.refs\fi
\chardef\wfile=\rfile\immediate\write\rfile{\noexpand\item{[\noexpand\hyperdef%
\noexpand\hypernoname{reference}{\the\refno}{\the\refno}]\ }%
\reflabeL{#1\hskip.31in}\pctsign}\global\advance\refno by1\findarg}
\def\findarg#1#{\begingroup\obeylines\newlinechar=`\^^M\pass@rg}
{\obeylines\gdef\pass@rg#1{\writ@line\relax #1^^M\hbox{}^^M}%
\gdef\writ@line#1^^M{\expandafter\toks0\expandafter{\striprel@x #1}%
\edef\next{\the\toks0}\ifx\next\em@rk\let\next=\endgroup\else\ifx\next\empty%
\else\immediate\write\wfile{\the\toks0}\fi\let\next=\writ@line\fi\next\relax}}
\def\striprel@x#1{} \def\em@rk{\hbox{}}
\def\lref{\begingroup\obeylines\lr@f}
\def\lr@f#1#2{\DefWarn#1\gdef#1{\let#1=\UNd@FiNeD\ref#1{#2}}\endgroup\unskip}
\def\semi{;\hfil\break}
\def\addref#1{\immediate\write\rfile{\noexpand\item{}#1}} 
\def\listrefs{\vfill\supereject\immediate\closeout\rfile\writestoppt
\baselineskip=\footskip\centerline{{\bf References}}\bigskip{\parindent=20pt%
\frenchspacing\escapechar=` \input \jobname.refs\vfill\eject}\nonfrenchspacing}
\def\startrefs#1{\immediate\openout\rfile=\jobname.refs\refno=#1}
\def\xref{\expandafter\xr@f}\def\xr@f[#1]{#1}
\def\refs#1{\count255=1[\r@fs #1{\hbox{}}]}
\def\r@fs#1{\ifx\UNd@FiNeD#1\message{reflabel \string#1 is undefined.}%
\nref#1{need to supply reference \string#1.}\fi%
\vphantom{\hphantom{#1}}{\let\hyperref=\relax\xdef\next{#1}}%
\ifx\next\em@rk\def\next{}%
\else\ifx\next#1\ifodd\count255\relax\xref#1\count255=0\fi%
\else#1\count255=1\fi\let\next=\r@fs\fi\next}
%

%
\newwrite\ffile\global\newcount\figno \global\figno=1
\def\fig{fig.~\hyperref{}{figure}{\the\figno}{\the\figno}\nfig}
\def\nfig#1{\DefWarn#1%
\xdef#1{fig.~\noexpand\hyperref{}{figure}{\the\figno}{\the\figno}}%
\writedef{#1\leftbracket fig.\noexpand~\xfig#1}%
\ifnum\figno=1\immediate\openout\ffile=\jobname.figs\fi\chardef\wfile=\ffile%
{\let\hyperref=\relax
\immediate\write\ffile{\noexpand\medskip\noexpand\item{Fig.\ %
\noexpand\hyperdef\noexpand\hypernoname{figure}{\the\figno}{\the\figno}. }
\reflabeL{#1\hskip.55in}\pctsign}}\global\advance\figno by1\findarg}
\def\xfig{\expandafter\xf@g}\def\xf@g fig.\penalty\@M\ {}
\def\figs#1{figs.~\f@gs #1{\hbox{}}}
\def\f@gs#1{{\let\hyperref=\relax\xdef\next{#1}}\ifx\next\em@rk\def\next{}\else
\ifx\next#1\xfig #1\else#1\fi\let\next=\f@gs\fi\next}
%
\def\figin{\epsfcheck\figin}\def\figins{\epsfcheck\figins}
\def\epsfcheck{\ifx\epsfbox\UnDeFiNeD
\message{(NO epsf.tex, FIGURES WILL BE IGNORED)}
\gdef\figin##1{\vskip2in}\gdef\figins##1{\hskip.5in}
\else\message{(FIGURES WILL BE INCLUDED)}%
\gdef\figin##1{##1}\gdef\figins##1{##1}\fi}
\def\DefWarn#1{}
\def\figinsert{\goodbreak\topinsert}
\def\ifig#1#2#3{\DefWarn#1\xdef#1{fig.~\the\figno}
\writedef{#1\leftbracket fig.\noexpand~\the\figno}%
\figinsert\figin{\centerline{#3}}
\smallskip
\leftskip=0pt \rightskip=0pt
\baselineskip12pt\noindent
{{\bf Fig.~\the\figno}\ \ninepoint #2}
\medskip
\global\advance\figno by1\par\endinsert}
\newwrite\lfile
{\escapechar-1\xdef\pctsign{\string\%}\xdef\leftbracket{\string\{}
\xdef\rightbracket{\string\}}\xdef\numbersign{\string\#}}
\def\writedefs{\immediate\openout\lfile=label.defs \def\writedef##1{%
{\let\hyperref=\relax\let\hyperdef=\relax\let\hypernoname=\relax
 \immediate\write\lfile{\string\def\string##1\rightbracket}}}}%
\def\writestop{\def\writestoppt{\immediate\write\lfile{\string\pageno
 \the\pageno\string\startrefs\leftbracket\the\refno\rightbracket
 \string\def\string\secsym\leftbracket\secsym\rightbracket
 \string\secno\the\secno\string\meqno\the\meqno}\immediate\closeout\lfile}}
\def\writestoppt{}\def\writedef#1{}

\def\seclab#1{\DefWarn#1%
\xdef #1{\noexpand\hyperref{}{section}{\the\secno}{\the\secno}}%
\writedef{#1\leftbracket#1}\wrlabeL{#1=#1}}
\def\subseclab#1{\DefWarn#1%
\xdef #1{\noexpand\hyperref{}{subsection}{\the\secno.\the\subsecno}%
{\the\secno.\the\subsecno}}\writedef{#1\leftbracket#1}\wrlabeL{#1=#1}}
\def\applab#1{\DefWarn#1%
\xdef #1{\noexpand\hyperref{}{appendix}{\secn@m}{\secn@m}}%
\writedef{#1\leftbracket#1}\wrlabeL{#1=#1}}
\newwrite\tfile \def\writetoca#1{}
\def\leaderfill{\leaders\hbox to 1em{\hss.\hss}\hfill}
\def\writetoc{\immediate\openout\tfile=\jobname.toc
   \def\writetoca##1{{\edef\next{\write\tfile{\noindent ##1
   \string\leaderfill{
   \string\hyperref{}{page}{\noexpand\number\pageno}%
   {\noexpand\number\pageno}} \par}}\next}}
}
\newread\ch@ckfile
\def\listtoc{\immediate\closeout\tfile\immediate\openin\ch@ckfile=\jobname.toc
\ifeof\ch@ckfile\message{no file \jobname.toc, no table of contents this pass}%
\else\closein\ch@ckfile\centerline{\bf Contents}\nobreak\medskip%
{\baselineskip=16pt\footnotefont\parskip=0pt\catcode`\@=11\input\jobname.toc
\catcode`\@=12\bigbreak\bigskip}\fi}
\catcode`\@=12 
\def\tenpoint{\def\rm{\fam0\tenrm}
\textfont0=\tenrm \scriptfont0=\sevenrm \scriptscriptfont0=\fiverm
\textfont1=\teni  \scriptfont1=\seveni  \scriptscriptfont1=\fivei
\textfont2=\tensy \scriptfont2=\sevensy \scriptscriptfont2=\fivesy
\textfont\itfam=\tenit \def\it{\fam\itfam\tenit}\def\footnotefont{\ninepoint}%
\textfont\bffam=\tenbf \def\bf{\fam\bffam\tenbf}\def\sl{\fam\slfam\tensl}\rm}
\font\ninerm=cmr9 \font\sixrm=cmr6 \font\ninei=cmmi9 \font\sixi=cmmi6
\font\ninesy=cmsy9 \font\sixsy=cmsy6 \font\ninebf=cmbx9
\font\nineit=cmti9 \font\ninesl=cmsl9 \skewchar\ninei='177
\skewchar\sixi='177 \skewchar\ninesy='60 \skewchar\sixsy='60
\def\ninepoint{\def\rm{\fam0\ninerm}
\textfont0=\ninerm \scriptfont0=\sixrm \scriptscriptfont0=\fiverm
\textfont1=\ninei \scriptfont1=\sixi \scriptscriptfont1=\fivei
\textfont2=\ninesy \scriptfont2=\sixsy \scriptscriptfont2=\fivesy
\textfont\itfam=\ninei \def\it{\fam\itfam\nineit}\def\sl{\fam\slfam\ninesl}%
\textfont\bffam=\ninebf \def\bf{\fam\bffam\ninebf}\rm}
%
\hyphenation{anom-aly anom-alies coun-ter-term coun-ter-terms}

\global\newcount\subsubsecno \global\subsubsecno=0
\def\subsubsec#1\par{\global\advance\subsubsecno by1%
{\toks0{#1}\message{(\the\secno\the\subsecno\the\subsubsecno. \the\toks0)}}%
\ifnum\lastpenalty>9000\else\bigbreak\fi
\noindent{\it\hyperdef\hypernoname{subsubsection}{\the\secno.\the\subsecno\the\subsubsecno}%
{\the\secno.\the\subsecno.\the\subsubsecno.} #1}
\par\nobreak\medskip\nobreak\noindent\ignorespaces}

\def\DefWarn#1{}
\def\tikzcaption#1#2{\DefWarn#1\xdef#1{Fig.~\the\figno}
\writedef{#1\leftbracket Fig.\noexpand~\the\figno}%
{
\smallskip
\leftskip=20pt \rightskip=20pt \baselineskip12pt\noindent
{{\bf Fig.~\the\figno}\ \ninepoint #2}
\bigskip
\global\advance\figno by1 \par}}

\def\ntoalpha#1{%
\ifcase#1%
@%
\or A\or B\or C\or D\or E\or F\or G\or H\or I
\fi
}

\global\newcount\appno \global\appno=1
\def\applab#1{\xdef #1{\ntoalpha\appno}\writedef{#1\leftbracket#1}\wrlabeL{#1=#1}
\global\advance\appno by1}

\def\preprint#1 #2\par{\rightline{\vbox{\baselineskip12pt\hbox{#1}\hbox{#2}}}\vskip2cm}
%
\def\title#1\par{\centerline{\bf #1}\nopagenumbers\pageno=0}
\def\author#1\par{\bigskip\bigskip\centerline{#1}}

\newcount\addressno

\def\email#1#2{\unskip$^#1$\footnote{\null}{\kern-\parindent \llap{$^#1$\hskip1pt}email: #2}}

\def\startcenter{%
  \par
  \begingroup
  \leftskip=0pt plus 1fil
  \rightskip=\leftskip
  \parindent=0pt
  \parfillskip=0pt
}
\def\stopcenter{\endgroup}

\def\address{\bigskip%
  \ifnum\the\addressno=0\else\stopcenter\endgroup\fi
  \advance\addressno by 1%
  \begingroup
  \startcenter
  \it
  \obeylines
  \addressAux
}
\def\addressAux#1{#1}

\def\abstract{\stopcenter\endgroup\bigskip\bigskip\noindent}

\def\Dsl{\,\raise.15ex\hbox{/}\mkern-13.5mu D} 
\def\dsl{\raise.15ex\hbox{/}\kern-.57em\partial}
 
\def\boxeqn#1{\vcenter{\vbox{\hrule\hbox{\vrule\kern3pt\vbox{\kern3pt
	\hbox{${\displaystyle #1}$}\kern3pt}\kern3pt\vrule}\hrule}}}


\def\a{\alpha}
\def\b{{\beta}}
\def\g{{\gamma}}
\def\d{{\delta}}

\def\l{\lambda}

\def\t{{\theta}}

\def\half{{1\over 2}}

\def\({\left(}
\def\){\right)}
\def\cF{{\cal F}}
\def\cW{{\cal W}}

\def\cA{{\cal A}}

\font\tenshuffle=shuffle10 \font\sevenshuffle=shuffle7 \font\fiveshuffle=shuffle7 at 5pt
\def\shuffle{{%
\def\Dshuffle{\mathbin{\hbox{\tenshuffle\char'001}}}%
\def\Sshuffle{\mathbin{\hbox{\sevenshuffle\char'001}}}%
\def\SSshuffle{\mathbin{\hbox{\fiveshuffle\char'001}}}%
\mathchoice{\Dshuffle}{\Dshuffle}{\Sshuffle}{\SSshuffle}}}


\def\qed{\hbox{\hskip 3pt
\vbox{\hrule\hbox to 7pt{\vrule height 7pt\hfill\vrule}
\hrule}}\hskip3pt}

\overfullrule=0pt\relax

\frenchspacing

\newread\instream \openin\instream= label.defs
\ifeof\instream \message{No labels in advance yet. Wait till next pass.}
\else \closein\instream \input label.defs
\fi
\writedefs

\def\arXiv:#1].{\hepthStrip#1 \nil}
\def\hepthStrip#1 #2\nil{\href{http://arxiv.org/abs/#1}{arXiv:#1 #2\unskip}].}

\input epsf
\input amssym

\font\frakfont=eufm10 at 10pt
\def\cK{{\cal K}}
\def\cX{{\cal X}}
\def\cyclic#1{{\rm cyclic}(#1)}

\def\ce{\mathord{\hbox{\frakfont e}}}
\def\cf{\mathord{\hbox{\frakfont f}}}
\def\cm{\mathord{\hbox{\frakfont M}}}

\preprint DAMTP--2015--69

\vskip-25pt\relax
\title Berends--Giele recursions and the BCJ duality in superspace and components

\author
Carlos R. Mafra\email{\star}{mafra@ias.edu}$^\dagger$ and
Oliver Schlotterer\email{\ddagger}{olivers@aei.mpg.de}

\address
$^\star$Institute for Advanced Study, School of Natural Sciences,
Einstein Drive, Princeton, NJ 08540, USA
\medskip
$^\dagger$DAMTP, University of Cambridge
Wilberforce Road, Cambridge, CB3 0WA, UK
\medskip
$^\ddagger$Max--Planck--Institut f\"ur Gravitationsphysik,
Albert--Einstein--Institut,
Am Muehlenberg, 14476 Potsdam, Germany

\abstract
The recursive method of Berends and Giele to compute tree-level gluon
amplitudes is revisited using the framework of ten-dimensional super Yang--Mills.
First, we prove that the pure spinor formula to compute SYM tree amplitudes derived
in 2010 reduces to the standard Berends--Giele formula from the 80s when restricted
to gluon amplitudes and additionally determine the fermionic completion. Second,
using BRST cohomology manipulations in superspace, alternative representations of
the component amplitudes are explored and the Bern--Carrasco--Johansson relations
among partial tree amplitudes are derived in a novel way. Finally, it is shown how
the supersymmetric components of manifestly local BCJ-satisfying tree-level
numerators can be computed in a recursive fashion.

\Date {October 2015}


\lref\BerendsME{
  F.A.~Berends and W.T.~Giele,
  ``Recursive Calculations for Processes with n Gluons,''
Nucl.\ Phys.\ B {\bf 306}, 759 (1988).
}
\lref\BerendsZN{
  F.A.~Berends and W.T.~Giele,
  ``Multiple Soft Gluon Radiation in Parton Processes,''
Nucl.\ Phys.\ B {\bf 313}, 595 (1989).
}
\lref\BerendsHF{
  F.A.~Berends, W.T.~Giele and H.~Kuijf,
  ``Exact and Approximate Expressions for Multi - Gluon Scattering,''
Nucl.\ Phys.\ B {\bf 333}, 120 (1990).
}

\lref\EOMBBs{
	C.R.~Mafra and O.~Schlotterer,
	``Multiparticle SYM equations of motion and pure spinor BRST blocks,''
	JHEP {\bf 1407}, 153 (2014).
	[arXiv:1404.4986 [hep-th]].
}

\lref\cohomology{
	C.R.~Mafra and O.~Schlotterer,
  	``Cohomology foundations of one-loop amplitudes in pure spinor superspace,''
	[arXiv:1408.3605 [hep-th]].
}

\lref\nptFT{
	C.R.~Mafra, O.~Schlotterer, S.~Stieberger and D.~Tsimpis,
	``A recursive method for SYM n-point tree amplitudes,''
	Phys.\ Rev.\ D {\bf 83}, 126012 (2011).
	[arXiv:1012.3981 [hep-th]].
}
\lref\nptTreeA{
	C.R.~Mafra, O.~Schlotterer and S.~Stieberger,
	``Complete N-Point Superstring Disk Amplitude I. Pure Spinor Computation,''
	Nucl.\ Phys.\ B {\bf 873}, 419 (2013).
	[arXiv:1106.2645 [hep-th]].
}

\lref\nptTreeB{
  	C.R.~Mafra, O.~Schlotterer and S.~Stieberger,
	``Complete N-Point Superstring Disk Amplitude II. Amplitude and Hypergeometric Function Structure,''
	Nucl.\ Phys.\ B {\bf 873}, 461 (2013).
	[arXiv:1106.2646 [hep-th]].
}

\lref\wittentwistor{
	E.Witten,
        ``Twistor-Like Transform In Ten-Dimensions''
        Nucl.Phys. B {\bf 266}, 245~(1986).
}
\lref\psf{
 	N.~Berkovits,
	``Super-Poincare covariant quantization of the superstring,''
	JHEP {\bf 0004}, 018 (2000)
	[arXiv:hep-th/0001035].
}
\lref\BCJ{
	Z.~Bern, J.J.M.~Carrasco and H.~Johansson,
	``New Relations for Gauge-Theory Amplitudes,''
	Phys.\ Rev.\ D {\bf 78}, 085011 (2008).
	[arXiv:0805.3993 [hep-ph]].
}

\lref\KKref{
	R.~Kleiss and H.~Kuijf,
	``Multi - Gluon Cross-sections and Five Jet Production at Hadron Colliders,''
	Nucl.\ Phys.\ B {\bf 312}, 616 (1989).
}
\lref\lanceKK{
	V.~Del Duca, L.J.~Dixon and F.~Maltoni,
	``New color decompositions for gauge amplitudes at tree and loop level,''
	Nucl.\ Phys.\ B {\bf 571}, 51 (2000).
	[hep-ph/9910563].
}
\lref\towards{
	C.R.~Mafra,
	``Towards Field Theory Amplitudes From the Cohomology of Pure Spinor Superspace,''
	JHEP {\bf 1011}, 096 (2010).
	[arXiv:1007.3639 [hep-th]].
}

\lref\PolicastroVT{
  G.~Policastro and D.~Tsimpis,
  ``$R^4$, purified,''
Class.\ Quant.\ Grav.\  {\bf 23}, 4753 (2006).
[hep-th/0603165].
}

\lref\HarnadBC{
  J.P.~Harnad and S.~Shnider,
  ``Constraints And Field Equations For Ten-dimensional Superyang-mills Theory,''
Commun.\ Math.\ Phys.\  {\bf 106}, 183 (1986).
}

\lref\MafraGIA{
 C.R.~Mafra and O.~Schlotterer,
  ``Solution to the nonlinear field equations of ten dimensional supersymmetric Yang-Mills theory,''
Phys.\ Rev.\ D {\bf 92}, no. 6, 066001 (2015).
[arXiv:1501.05562 [hep-th]].
}

\lref\anomaly{
	N.~Berkovits and C.R.~Mafra,
	``Some Superstring Amplitude Computations with the Non-Minimal Pure Spinor Formalism,''
	JHEP {\bf 0611}, 079 (2006).
	[hep-th/0607187].
}

\lref\MafraKH{
	C.R.~Mafra and O.~Schlotterer,
  	``The Structure of n-Point One-Loop Open Superstring Amplitudes,''
	JHEP {\bf 1408}, 099 (2014).
	[arXiv:1203.6215 [hep-th]].
}

\lref\MafraKJ{
	C.R.~Mafra, O.~Schlotterer and S.~Stieberger,
  	``Explicit BCJ Numerators from Pure Spinors,''
	JHEP {\bf 1107}, 092 (2011).
	[arXiv:1104.5224 [hep-th]].
}

\lref\MafraGJA{
	C.R.~Mafra and O.~Schlotterer,
  	``Towards one-loop SYM amplitudes from the pure spinor BRST cohomology,''
	Fortsch.\ Phys.\  {\bf 63}, no. 2, 105 (2015).
	[arXiv:1410.0668 [hep-th]].
}

\lref\MafraMJA{
	C.R.~Mafra and O.~Schlotterer,
	``Two-loop five-point amplitudes of super Yang-Mills and supergravity in pure spinor superspace,''
	JHEP {\bf 1510}, 124 (2015).
	[arXiv:1505.02746 [hep-th]].
}

\lref\MafraPN{
	C.R.~Mafra,
  	``PSS: A FORM Program to Evaluate Pure Spinor Superspace Expressions,''
	[arXiv:1007.4999 [hep-th]].
}

\lref\WWW{
	C.R.~Mafra and O.~Schlotterer,
	``PSS: From pure spinor superspace to components,''
	{\tt http://www.damtp.cam.ac.uk/user/crm66/SYM/pss.html}.
}

\lref\SiegelYI{
  W.~Siegel,
  ``Superfields in Higher Dimensional Space-time,''
Phys.\ Lett.\ B {\bf 80}, 220 (1979).
}

\lref\BernUE{
  Z.~Bern, J.~J.~M.~Carrasco and H.~Johansson,
  ``Perturbative Quantum Gravity as a Double Copy of Gauge Theory,''
Phys.\ Rev.\ Lett.\  {\bf 105}, 061602 (2010).
[arXiv:1004.0476 [hep-th]].
}

\lref\BernYG{
  Z.~Bern, T.~Dennen, Y.~t.~Huang and M.~Kiermaier,
  ``Gravity as the Square of Gauge Theory,''
Phys.\ Rev.\ D {\bf 82}, 065003 (2010).
[arXiv:1004.0693 [hep-th]].
}

\lref\BerkovitsRB{
  N.~Berkovits,
  ``Covariant quantization of the superparticle using pure spinors,''
JHEP {\bf 0109}, 016 (2001).
[hep-th/0105050].
}

\lref\companion{
  S.~Lee, C.R.~Mafra and O.~Schlotterer,
  ``Non-linear gauge transformations in $D=10$ SYM and the BCJ duality,''
[arXiv:1510.08843 [hep-th]].
}

\lref\OprisaWU{
  D.~Oprisa and S.~Stieberger,
  ``Six gluon open superstring disk amplitude, multiple hypergeometric series and Euler-Zagier sums,''
[hep-th/0509042].
}

\lref\StiebergerTE{
  S.~Stieberger and T.~R.~Taylor,
  ``Multi-Gluon Scattering in Open Superstring Theory,''
Phys.\ Rev.\ D {\bf 74}, 126007 (2006).
[hep-th/0609175].
}

\lref\BroedelTTA{
  J.~Broedel, O.~Schlotterer and S.~Stieberger,
  ``Polylogarithms, Multiple Zeta Values and Superstring Amplitudes,''
Fortsch.\ Phys.\  {\bf 61}, 812 (2013).
[arXiv:1304.7267 [hep-th]].
}

\lref\BjerrumBohrRD{
  N.~E.~J.~Bjerrum-Bohr, P.~H.~Damgaard and P.~Vanhove,
  ``Minimal Basis for Gauge Theory Amplitudes,''
Phys.\ Rev.\ Lett.\  {\bf 103}, 161602 (2009).
[arXiv:0907.1425 [hep-th]].
}

\lref\StiebergerHQ{
  S.~Stieberger,
  ``Open \& Closed vs. Pure Open String Disk Amplitudes,''
[arXiv:0907.2211 [hep-th]].
}

\lref\ChenJXA{
  Y.~X.~Chen, Y.~J.~Du and B.~Feng,
  ``A Proof of the Explicit Minimal-basis Expansion of Tree Amplitudes in Gauge Field Theory,''
JHEP {\bf 1102}, 112 (2011).
[arXiv:1101.0009 [hep-th]].
}
\lref\BGschocker{
M. Schocker, 
``Lie elements and Knuth relations,'' Canad. J. Math. {\bf 56} (2004), 871-882.
[math/0209327].
}

\lref\SelivanovHN{
  K.G.~Selivanov,
  ``On tree form-factors in (supersymmetric) Yang-Mills theory,''
Commun.\ Math.\ Phys.\  {\bf 208}, 671 (2000).
[hep-th/9809046].
}

\lref\ree{
	R. Ree,
	``Lie elements and an algebra associated with shuffles'',
	Ann. Math. {\bf 68}, No. 2 (1958), 210--220.
}

\lref\BrinkBC{
  L.~Brink, J.H.~Schwarz and J.~Scherk,
  ``Supersymmetric Yang-Mills Theories,''
Nucl.\ Phys.\ B {\bf 121}, 77 (1977).
}

\lref\bigHowe{
  P.S.~Howe,
  ``Pure Spinors Lines In Superspace And Ten-Dimensional Supersymmetric
  Theories,''
  Phys.\ Lett.\  B {\bf 258}, 141 (1991)
  [Addendum-ibid.\  B {\bf 259}, 511 (1991)].
\semi
  P.S.~Howe,
  ``Pure Spinors, Function Superspaces And Supergravity Theories In
  Ten-Dimensions And Eleven-Dimensions,''
  Phys.\ Lett.\  B {\bf 273}, 90 (1991).
}
\lref\twoloopids{
	C.R.~Mafra,
  	``Pure Spinor Superspace Identities for Massless Four-point Kinematic Factors,''
	JHEP {\bf 0804}, 093 (2008).
	[arXiv:0801.0580 [hep-th]].
}
\lref\fivetree{
	C.R.~Mafra,
	``Simplifying the Tree-level Superstring Massless Five-point Amplitude,''
	JHEP {\bf 1001}, 007 (2010).
	[arXiv:0909.5206 [hep-th]].
}

\lref\sixtree{
	C.R.~Mafra, O.~Schlotterer, S.~Stieberger and D.~Tsimpis,
  	``Six Open String Disk Amplitude in Pure Spinor Superspace,''
	Nucl.\ Phys.\ B {\bf 846}, 359 (2011).
	[arXiv:1011.0994 [hep-th]].
}

\lref\Selivanov{
A.A.~Rosly and K.G.~Selivanov,
  ``On amplitudes in selfdual sector of Yang-Mills theory,''
Phys.\ Lett.\ B {\bf 399}, 135 (1997).
[hep-th/9611101].
\semi
	A.A.~Rosly and K.G.~Selivanov,
  	``Gravitational SD perturbiner,''
	[hep-th/9710196].
\semi
  K.G.~Selivanov,
  ``Postclassicism in tree amplitudes,''
[hep-th/9905128].
}
\lref\Bardeen{
  W.A.~Bardeen,
  ``Selfdual Yang-Mills theory, integrability and multiparton amplitudes,''
Prog.\ Theor.\ Phys.\ Suppl.\  {\bf 123}, 1 (1996).
}
\lref\JJreview{
  J.~J.~M.~Carrasco,
  ``Gauge and Gravity Amplitude Relations,''
[arXiv:1506.00974 [hep-th]].
}

\lref\DaviesVT{
  S.~Davies,
  ``One-Loop QCD and Higgs to Partons Processes Using Six-Dimensional Helicity and Generalized Unitarity,''
Phys.\ Rev.\ D {\bf 84}, 094016 (2011).
[arXiv:1108.0398 [hep-ph]].
}

\lref\TolottiCAA{
  M.~Tolotti and S.~Weinzierl,
  ``Construction of an effective Yang-Mills Lagrangian with manifest BCJ duality,''
JHEP {\bf 1307}, 111 (2013).
[arXiv:1306.2975 [hep-th]].
}

\lref\BadgerGXA{
  S.~Badger, H.~Frellesvig and Y.~Zhang,
  ``A Two-Loop Five-Gluon Helicity Amplitude in QCD,''
JHEP {\bf 1312}, 045 (2013).
[arXiv:1310.1051 [hep-ph]].
}

\lref\EllisIR{
  R.~K.~Ellis, W.~T.~Giele, Z.~Kunszt and K.~Melnikov,
  ``Masses, fermions and generalized $D$-dimensional unitarity,''
Nucl.\ Phys.\ B {\bf 822}, 270 (2009).
[arXiv:0806.3467 [hep-ph]].
}

\listtoc
\writetoc
\filbreak

\newsec Introduction

Ten-dimensional super Yang--Mills (SYM) provides a simplified description of maximally 
supersymmetric gauge theories \wittentwistor. On the one hand, its spectrum comprises 
just a gluon and a gluino which automatically cover the scalars in lower-dimensional 
formulations \BrinkBC. On the other hand, pure spinors allow to formulate the on-shell 
conditions as a cohomology problem \bigHowe, and the BRST operator in the associated 
pure spinor superspace powerfully embodies gauge invariance and supersymmetry \psf.
This framework naturally appears in the manifestly super Poincar\'e-covariant quantization of the superstring \psf.

Using a confluence of string-theory techniques and field-theory intuition,
scattering amplitudes in ten-dimensional SYM have been compactly represented
in pure spinor superspace \refs{\nptFT, \MafraGJA,
\MafraMJA}. This construction crucially rests on the notion of multiparticle
superfields \EOMBBs\ which were motivated by superstring computations
\refs{\twoloopids \fivetree \sixtree \nptTreeA {--} \MafraKH}. Multiparticle
superfields collect the contributions of tree-level subdiagrams at arbitrary
multiplicity and can be flexibly attached to multiloop diagrams, see \MafraMJA\
for a two-loop application.

In a companion paper \companion, the construction of multiparticle superfields and their 
expansion in the Grassmann variable $\theta^\alpha$ of pure spinor superspace have 
been tremendously simplified. In the following, we will revisit tree-level amplitudes in the 
light of the new theta-expansions and in particular:
\medskip
\item{$\bullet$} recover and supersymmetrize the Berends--Giele recursion for gluonic tree amplitudes
\item{$\bullet$} present a simplified component realization of the BCJ color-kinematics duality, 
along with a new superspace proof for the closely related BCJ relations.

\subsec Summary of results on the Berends--Giele recursion
\par\subseclab\onetwo

\noindent
The theta-expansions of ten-dimensional multiparticle superfields
have recently \companion\ been simplified using supersymmetric Berends--Giele
currents which generalize the gluonic currents defined by Berends and
Giele \BerendsME. Using these simplified expansions, the pure spinor superspace formula
to compute ten-dimensional color-ordered SYM amplitudes at tree level \nptFT,
\eqn\PSformula{
A^{\rm SYM}(1,2, \ldots,p,p+1) = \langle E_{12 \ldots p} M_{p+1}\rangle \ ,
}
will be explicitly evaluated in components and shown to be
\eqn\BGfullIntro{
A^{\rm SYM}(1,2, \ldots,p,p+1) =
s_{12 \ldots p} (\ce_{12 \ldots p}\cdot \ce_{p+1}) + k^m_{12 \ldots p} (\cX_{12 \ldots p}\g_m \cX_{p+1})\,.
}
The superfields $E_{12 \ldots p}$ and $M_{p+1}$ will be introduced in section \SYMsec, and the square of the
momentum $k^m_{12 \ldots p}\equiv k_1^m+k_2^m+\cdots+k_p^m$ is denoted by $s_{12\ldots p}$. 
Moreover, $\ce^m_{12\ldots p}$ and ${\cal X}^\alpha_{12\ldots p}$ in \BGfullIntro\ denote
the component Berends--Giele currents which
depend on the gluon and gluino polarizations $e_i^m, \chi_i^\a$
as well as light-like momenta $k_i^m$ associated with legs $i=1,2,\ldots ,p$.
Finally, $m=0, \ldots,9$ and $\a=1, \ldots,16$ are vector and Weyl-spinor indices
of the Lorentz group $SO(1,9)$.

After setting the fermionic fields to zero, the first term in \BGfullIntro\ will be
shown to reproduce the gluonic Berends--Giele formula \BerendsME ,
\eqn\BGYM{
A^{\rm YM}(1,2,\ldots, p+1) = s_{12 \ldots p}(J_{12 \ldots p}\cdot J_{p+1}) \ ,
}
making \BGfullIntro\ its supersymmetric generalization for ten-dimensional SYM.

Furthermore, the same Berends--Giele currents $\ce_{12\ldots p}^m$ and
${\cal X}_{12\ldots p}^{\alpha}$ together with a
field-strength companion $\cf_{12\ldots p}^{mn}$ will be shown to yield economic
and manifestly cyclic representations of SYM amplitudes such as
\eqnn\explone
$$\eqalignno{
A^{\rm SYM}(1,2,3,4,5) &=  {1\over 2} (\ce^m_{12} \cf^{mn}_{34} \ce_{5}^n
+\ce^m_{34} \cf^{mn}_{5} \ce_{12}^n
+\ce^m_{5} \cf^{mn}_{12} \ce_{34}^n)  &\explone\cr
&+({\cal X}_{12} \g_{m} {\cal X}_{5}) \ce^m_{34}
+({\cal X}_{34} \g_{m} {\cal X}_{12}) \ce^m_{5}
+({\cal X}_{5} \g_{m} {\cal X}_{34}) \ce^m_{12} + \cyclic{12345}  \ ,
}
$$
streamlining the earlier approach in \BerendsHF\ based on the above $J^m_{12\ldots p}$.

Using the generating series of supersymmetric Berends--Giele currents
discussed in \refs{\MafraGIA,\companion}, it will be shown that
the generating series of ten-dimensional SYM tree-level amplitudes
takes a very simple form,
\eqn\simpleGen{
{\rm Tr} \Big( {1\over 4} \Bbb F_{mn} \Bbb F^{mn}
+ (\Bbb W  \gamma^m \nabla_m \Bbb W) \Big) \, \Big|_{\theta=0} =
\sum_{n=3}^{\infty} {n-2\over n}\!\!\!\!\! \sum_{i_1,i_2,\ldots,i_n}
\!\!\!\!\!{\rm Tr}(t^{i_1} t^{i_2} \ldots t^{i_n}) A^{{\rm SYM}}(i_1,i_2,\ldots,i_n)\,.
}
Note that the left-hand side of \simpleGen\ matches the ten-dimensional
SYM Lagrangian evaluated on the generating series $\Bbb F^{mn}(x,\t=0)$ and
$\Bbb W^\a(x,\t=0)$ defined below.

\subsec Summary of results on the BCJ duality
\par\subseclab\oneeight

\noindent
The virtue of the simplified theta-expansions in \companion\ can be reconciled with a 
manifestation of the duality between color and kinematics due to Bern, Carrasco and 
Johansson (BCJ) \BCJ\ (see \JJreview\ for a review).
A concrete tree-level realization of the BCJ duality was given 
in \MafraKJ\ at any multiplicity, based on local numerators in pure spinor superspace.
The components are accessible through the zero-mode treatment in \MafraPN,
but we will present a significantly accelerated approach where the zero-mode manipulations
are trivialized.

The BCJ duality immediately led to the powerful prediction that only $(n-3)!$ 
permutations of SYM tree-level subamplitudes \BGfullIntro\ are linearly independent \BCJ.
This basis dimension was later derived from the
monodromy properties of the string worldsheet \refs{\BjerrumBohrRD, \StiebergerHQ},
by the field-theory limit of the $n$-point superstring disk
amplitude \refs{\nptTreeA,\nptTreeB} and by BCFW on-shell recursions in field theory \ChenJXA.
In addition to these proofs,
the following explicit BCJ relations among color-ordered amplitudes
will be obtained from pure spinor cohomology arguments,
\eqn\BCJrelIntro{
\sum_{i=1}^{|A|} \sum_{j=1}^{|B|} (-1)^{i-j} s_{a_i b_j}
A^{\rm SYM}(
(a_1 \ldots a_{i-1} \shuffle a_{|A|} \ldots a_{i+1}),a_i,b_j,
(b_{j-1}\ldots   b_1 \shuffle b_{j+1} \ldots b_{|B|})
,n) = 0\,,
}
where the words $A=a_1 a_2\ldots a_{|A|}$ and $B=b_1b_2\ldots b_{|B|}$ have total
length $|A|+|B|=n-1$. The shuffle product $\shuffle$ is defined recursively
as 
\eqn\Shrecurs{
\emptyset\shuffle A = A\shuffle\emptyset = A,\qquad
A\shuffle B \equiv a_1(a_2 \ldots a_{|A|} \shuffle B) + b_1(b_2 \ldots b_{|B|}
\shuffle A)\,,
}
where $\emptyset$ denotes the case when no ``letter'' is present.

\newsec Review

\subsec Berends--Giele recursion relations
\par\subseclab\twoone

\noindent
In the 80s, Berends and Giele proposed a recursive method to compute color-ordered
gluon amplitudes at tree level using multiparticle currents $J^m_P$
defined\foot{The
original definition of $J^m_P$ in \BerendsME\ contains the factor
$1/k_P^2$ instead of $1/s_P$ as adopted here. An overall
factor of $\half$ in \cubicV\ and \quarticV\ 
compensates this difference.} as \BerendsME
\eqn\BGrecur{
J^m_i \equiv e_i^m\,,\qquad
s_P J^m_P \equiv \sum_{XY=P} [J_X, J_Y]^m + \sum_{XYZ=P}\{J_X,J_Y,J_Z\}^m \ ,
}
where $e^m_i$ denotes the polarization vector of a single-particle gluon,
$P=12\ldots p$ encompasses several external particles, and the Mandelstam invariants are
\eqn\manddef{
s_P\equiv \half k_P^2 \ , \ \ \ \ \ \ \ k_P^m\equiv k^m_1 + k^m_2 + \cdots + k^m_p\,.
}
The notation $\sum_{XY=P}$ in \BGrecur\ instructs to deconcatenate $P = 12\ldots p$ into 
non-empty words $X = 12\ldots j$ and $Y=j+1\ldots p$
with $j=1,2,\ldots,p-1$ and the obvious generalization to $\sum_{XYZ=P}$. 
The brackets $[\cdot,\cdot]^m$ and $\{\cdot, \cdot, \cdot \}^m$ are given by stripping off
one gluon field (with vector index $m$) from the cubic and
quartic vertices of the Yang--Mills Lagrangian,
\eqnn\cubicV
\eqnn\quarticV
$$\eqalignno{
[J_X,J_Y]^m &\equiv  (k_Y\cdot J_X)J_Y^m + \half k_X^m (J_X\cdot J_Y) -
(X\leftrightarrow Y) &\cubicV\cr
\{J_X,J_Y,J_Z\}^m & \equiv  (J_X\cdot J_Z)J_Y^m - \half(J_X\cdot J_Y)J_Z^m -\half
(J_Y\cdot J_Z)J_X^m \,. &\quarticV
}$$
The {\it Berends--Giele currents} $J^m_P$ are conserved \BerendsME\ and satisfy
certain symmetries
\BerendsZN,
\eqn\BGsym{
k^m_P J^m_P = 0\,,\qquad
J^m_{A\shuffle B} = 0, \quad \forall A,B\neq\emptyset \ .
}
The purely gluonic amplitudes are then computed as \BerendsME\ 
\eqn\BGformula{
A^{\rm YM}(1,2, \ldots,p,p+1) = s_{12 \ldots p}J^m_{12 \ldots p}J^m_{p+1} \ .
}
For example, the Berends--Giele current of multiplicity two following from
\BGrecur\ is
\eqn\rankTwoJ{
s_{12} J^m_{12} = e^m_2 (e_1\cdot k_2) - e^m_1 (e_2\cdot k_1) +
\half(k_1^m-k_2^m)(e_1\cdot e_2)
}
and leads to the well-known three-point amplitude
\eqn\BGthreept{
A^{\rm YM}(1,2,3) = s_{12}J^m_{12}J^m_3 = (e_1\cdot e_2)(k_1\cdot e_3)+{\rm cyclic}(123)\,.
}
Note that the Berends--Giele formula \BGformula\ as presented in \BerendsME\ is not
supersymmetric, it computes purely gluonic amplitudes.

\subsec Super Yang--Mills superfields in ten dimensions
\par\subseclab\twotwo

\noindent SYM in ten dimensions admits a super-Poincare-invariant
description in terms of four types of superfields: the
spinor potential $\Bbb
A_\a(x,\t)$, the vector potential $\Bbb A^m(x,\t)$ and their associated
field-strengths $\Bbb W^\a(x,\t)$, $\Bbb F^{mn}(x,\t)$. They
satisfy the following non-linear field
equations\foot{Our convention for (anti)symmetrizing indices does not include ${1\over 2}$, e.g.~$\partial^{[m} \gamma^{n]}\!=\!\partial^{m} \gamma^{n}\!-\! \partial^{n} \gamma^{m}$.} \wittentwistor,
\eqnn\SYMeom
$$\eqalignno{
\{D_{(\alpha},\Bbb A_{\beta) } \}&=\gamma^{m}_{\alpha\beta}\Bbb A_m+\{\Bbb A_\alpha, \Bbb A_\beta\} &\SYMeom\cr
[D_{\alpha},\Bbb A_m]&=[\partial_m,\Bbb A_\alpha]+(\gamma_m \Bbb W)_\alpha+[\Bbb A_\alpha,\Bbb A_m] \cr
\{D_\alpha,\Bbb W^\beta\}&={1\over4}(\gamma^{mn})_\alpha^{\phantom{\alpha}\beta}\Bbb F_{mn}+\{\Bbb A_\alpha,\Bbb W^\beta\} \cr
[D_\alpha, \Bbb F^{mn}]&= [\partial^{[m} , (\Bbb W \gamma^{n]})_\alpha] -  [\Bbb A^{[m}, (\Bbb W \gamma^{n]})_\alpha] +[\Bbb A_\alpha,\Bbb
F^{mn}]\,.
}$$
For later convenience,
we use the notation where $\Bbb K$ refers to any element of the set containing
these superfields,
\eqn\BbbKs{
\Bbb K \in \{\Bbb A_\a, \Bbb A_m,
\Bbb W^\a, \Bbb F^{mn}\}\,.
}
In the context of scattering amplitudes or vertex operators of the
superstring \psf, one discards the quadratic terms from
\SYMeom\ to obtain the linearized superfields of ten-dimensional
SYM $K_i \in \{A^i_\a, A^i_m, W^\a_i, F_i^{mn}\}$
satisfying
\eqn\SYM{
\eqalign{
\{ D_{(\a} , A^i_{\b)} \}  &= \g^m_{\a\b} A^i_m \ ,\cr
[ D_\a, A^i_m ] &= (\g_m W_i)_\a + [\partial_m ,A^i_\a] \ ,\cr
}
\qquad\eqalign{
\{D_\a ,W^{\b}_i \} &= {1\over 4}(\g^{mn})_\a{}^\b F^i_{mn}\cr
 [ D_\a , F^i_{mn} ] &= [\partial_{[m}, (\g_{n]} W_i)_\a]\, .
}}
They describe a single gluon and/or gluino which furnishes the $i^{\rm th}$
leg in the amplitude.

In pursuing compact expressions for superstring scattering amplitudes one is led to
a natural multiparticle generalization of the above description, where
the single-particle labels are replaced by ``words'' $P=123 \ldots p$.
In particular, amplitudes can be compactly written in terms of
non-local\foot{A discussion of {\it local}
multiparticle superfields $K_P$ can be found in \refs{\companion,\EOMBBs}.}
superfields called Berends--Giele currents
$\cK_P \in \{\cA^P_\a, \cA^P_m, \cW^\a_P, \cF_P^{mn}\}$ encompassing several 
legs $1,2,\ldots,p$ in an amplitude. They are recursively 
constructed from linearized superfields in \SYM, and the original
expressions in \EOMBBs\ are related
to simplified representations in \companion\ via non-linear gauge
transformations. This gauge freedom
affects the generating series $\Bbb K \in \{\Bbb A_\a, \Bbb A_m, \Bbb W^\a, \Bbb F^{mn}\}$ of 
Berends--Giele currents
\eqn\Aser{
\Bbb K = \sum_i {\cal K}_i t^i + \sum_{i,j} {\cal K}_{ij} t^i t^j
 + \sum_{i,j,k} {\cal K}_{ijk} t^i t^j t^k + \cdots  \,,
}
where $t^i$ are generators of a non-abelian gauge group. The generating
series in \Aser\ were shown in \MafraGIA\ to solve the
non-linear field equations\foot{It should be pointed out that the notion of a generating series which
solves the field equations
and gives rise to tree amplitudes corresponds to
the ``perturbiner'' formalism \Selivanov. This approach has been
applied to the self-dual sector of Yang--Mills theory and led
to a generating series of MHV amplitudes, see \SelivanovHN\ for a supersymmetric extension. However, the generic
Yang--Mills amplitudes have never been obtained this way (see also \Bardeen).
We thank Nima Arkani-Hamed for pointing out these references.
} \SYMeom\ by the properties of the constituent
Berends--Giele currents $\cK_P \in \{\cA^P_\a, \cA^P_m, \cW^\a_P, \cF_P^{mn}\}$.

\subsubsec Simplifying component expansions with superfield gauge transformations

The aforementioned gauge freedom of the generating series \Aser\ allows to tune the
theta-expansion of the  multiparticle supersymmetric Berends--Giele currents
such that \companion
\eqn\thetaEXP{
{\cal A}^P_{\alpha}(x,\t)= \Big({1\over 2}(\t\g_m)_\alpha \ce^m_P
+{1\over 3}(\t\g^m)_\alpha (\t\g_m{\cal X}_P)
- {1\over 32} (\g^p \t)_\alpha (\t \g_{mnp} \t) \cf_{P}^{mn}+ \! \ldots \! \Big) e^{k_P
\cdot x}
}
takes the same form as the linearized superfield $A^i_\alpha$ subject to \SYM\ \refs{\HarnadBC, \PolicastroVT},
\eqn\linEXP{
A^i_{\alpha}(x,\t)= \Big({1\over 2}(\t\g_m)_\alpha e^m_i
+{1\over 3}(\t\g^m)_\alpha (\t\g_m \chi_i)
- {1\over 32} (\g^p \t)_\alpha (\t \g_{mnp} \t)f_{i}^{mn}+ \! \ldots \! \Big) e^{k_i
\cdot x}\,.
}
The components $\ce^m_P, {\cal X}^\alpha_P, \cf_{P}^{mn}$ depend on the momenta $k^m_i$,
polarizations $e^m_i$ and wavefunctions $\chi^\alpha_i$ of the gluons and gluinos encompassed in the
multiparticle label $P=12\ldots p$ and can be obtained from the recursions \companion
\eqn\reczero{
\ce^m_P = {1 \over s_P} \sum_{XY=P} \ce^{m}_{[X,Y]} \ ,
\quad {\cal X}^\alpha_P = {1 \over s_P} \sum_{XY=P} {\cal X}^{\alpha}_{[X,Y]} \ ,
}
where $\ce^m_i \equiv e^m_i$ and $\cX^\a_i\equiv \chi^\a_i$ for a single-particle
label as well as
\eqnn\recone
\eqnn\rectwo
$$\eqalignno{
\ce^{m}_{[X,Y]}  &\equiv - {1 \over 2 }  \bigl[ \ce_{X}^m (k^X\cdot  \ce^{Y})
+ \ce^{X}_n  \cf_Y^{mn}
- ( {\cal X}^{X}\g^m {\cal X}^Y)
- (X \leftrightarrow Y)\bigr] &\recone\cr
{\cal X}^{\alpha}_{[X,Y]} &\equiv 
{1 \over 2 } ( k^p_{X} + k^p_{Y} ) \gamma_p^{\alpha \beta}
\big[ \ce_X^m ( \g_m {\cal X}_Y)_\beta - \ce_Y^m (\g_m \cX_X)_\beta\big]  \ .&\rectwo\cr
}$$
The non-linear component field-strength is given by
\eqnn\recthree
$$\eqalignno{
\cf^{mn}_P &\equiv k_P^m \ce_P^n - k_P^n \ce_P^m
- \sum_{XY=P}\!\!\big( \ce_X^m \ce_Y^n - \ce_X^n \ce_Y^m \big) 
&\recthree
}$$
and generalizes the single-particle
instance $\cf^{mn}_i \equiv f^{mn}_i = k_i^m e_i^n - k_i^n e_i^m$ in \linEXP.

The expressions in \recone, \rectwo\ and \recthree\ are obtained from the
theta-independent terms of the superfields $\cA^m_P,\cW^\a_P,\cF^{mn}_P$
evaluated at $x=0$ \companion,
\eqn\zerocomp{
\ce^{m}_P \equiv \cA^m_P(0,0) \ , \ \ \ \ \ \ {\cal X}_P^\alpha \equiv \cW^\a_P(0,0) \ , \ \ \ \ \ \
\cf^{mn}_P \equiv \cF_P^{mn}(0,0) \ ,
}
in the same way as $e^m_i,\chi^\alpha_i$ and $f^{mn}_i$ stem from the
linearized superfields $A^m_i,W^\alpha_i,F^{mn}_i$. Accordingly, the
recursions in \reczero\ to \rectwo\ for $\ce^m_P$ and ${\cal X}_P^\alpha$
descend from the recursive construction of superspace Berends--Giele
currents $\cA^m_P,\cW^\a_P,\cF^{mn}_P$ described in \companion.

Note that the transversality of the gluon and the Dirac equation of the gluino
propagate as follows to the multiparticle level,
\eqn\eoms{
(k_P \cdot \ce_P) = 0 \ , \qquad k^P_m (\g^m {\cal X}_P)_\a
=
\sum_{XY=P}\big[ \ce_X^m (\g_m \cX_Y)_\a -  \ce_Y^m (\g_m \cX_X)_\a \big],
}
where transversality of $\ce_P^m$ is a peculiarity of the Lorentz gauge chosen in
the derivation of the corresponding superspace Berends--Giele current
$\cA^m_P(x,\t)$ \companion.

\subsec The pure spinor superspace formula for SYM tree amplitudes
\par\subseclab\SYMsec

\noindent
Tree-level amplitudes in ten-dimensional SYM have been constructed in \nptFT\ from cohomology
methods in pure spinor superspace \psf. Inspired by OPEs in string theory,
the BRST-invariant superspace expression
\eqn\AYM{
A^{\rm SYM}(1,2, \ldots,p,p+1) = \langle E_{12 \ldots p}M_{p+1}\rangle \equiv
\!\!\!\sum_{XY=12\ldots p}\!\!\! \langle  M_X M_Y M_{p+1}\rangle
}
with the pole structure of a color-ordered $(p+1)$-point amplitude has
been proposed and shown to
reproduce known component expressions for various combinations of gluons and
gluinos. BRST invariance of the superfields implies gauge-invariant and
supersymmetric components. In \AYM\
the bracket $\langle \ldots \rangle$ instructs to pick up terms of order $\l^3
\theta^5$ of the enclosed superfields \psf, and
the following shorthand has been used
\eqn\Mdef{
M_P \equiv \l^\a \cA_\a^P(x,\t)
}
for contractions of the pure spinor $\lambda^\alpha$. At this point, 
we make use of the gauge choice in \companion\ where the
theta-expansion \thetaEXP\ of the multiparticle superfield mimics the
single-particle counterpart \linEXP.  In this way, the same $\l^3\t^5$ correlators
listed on appendix A of \anomaly\ govern both the three-point amplitude
\eqn\PSthreept{
A^{\rm SYM}(1,2,3) = \langle M_1 M_2 M_3\rangle = \half \ce^m_1 \cf^{mn}_2 \ce^{n}_3 +
(\cX_1\g_m\cX_2)\ce^m_3 + \cyclic{123}\, 
}
and a generic multiparticle constituent of the $n$-point amplitudes \AYM,
\eqn\compon{
\langle M_X M_Y M_Z \rangle = \half\ce^m_X\,  \cf^{mn}_Y \ce^n_Z
+ (\cX_X \g_m \cX_Y) \ce_Z^m + \cyclic{XYZ} 
\equiv \cm_{X,Y,Z} \,.
}
This makes the gluon and gluino components of an arbitrary $n$-point tree amplitude
easily accessible through the recursion \reczero\ to \recthree\ for the components 
$\ce^m_P, {\cal X}^\alpha_P$ and $\cf^{mn}_P$. Using the component field-strength
\recthree, it follows that the gluonic three-point amplitudes of
the Berends--Giele and pure spinor formul{\ae} match.
In the following section, we will demonstrate that the same is
true for an arbitrary number of external legs.

\newsec The supersymmetric completion of the Berends--Giele formula

In this section, the pure spinor superspace formula for
ten-dimensional SYM tree amplitudes \AYM\
will be shown to reduce {\it ipsis litteris} to the
Berends--Giele formula \BGformula\ when restricted to its gluonic expansion.
Given the supersymmetry of the pure spinor approach, we will use it to
derive the supersymmetric completion of the Berends--Giele formula.

\subsec Bosonic Berends--Giele current from superfields
\par\subseclab\threeone

\noindent In a first step, the lowest components $\ce_P^m$ in the superfield
\thetaEXP\ are demonstrated to reproduce the bosonic Berends--Giele currents in
\BGrecur\ once the fermions are decoupled, i.e.
\eqn\repcurrent{
\ce^m_P \, \big|_{\chi_j = 0} = J^m_P \ .
}
Plugging the field-strength $\cf_P^{mn}$ \recthree\ into the recursive
definition of $\ce_P^m$ \reczero\ leads to
\eqnn\regroup
$$\eqalignno{
2s_P \ce^m_P  &= - \sum_{XY=P}
\bigl[ 2 \ce_{X}^m (k_X\cdot  \ce_{Y}) + k^{m}_Y  (\ce_X\cdot \ce_Y)
- ({\cal X}_X \g^m {\cal X}_Y)  - (X
\leftrightarrow Y)\bigr] \cr
&{}\ \ \ \ \ \ \ +  \sum_{XYZ=P} \bigl[2 (\ce_X\cdot \ce_Z)\ce_Y^m
- (\ce_X\cdot \ce_Y)\ce_Z^m - (\ce_Y\cdot \ce_Z)\ce_X^m
\bigr] \ .&\regroup\cr
}$$
In absence of fermions, $\chi^\alpha_j = 0$, the first line \regroup\ yields the
contribution of the cubic vertex \cubicV\ to the Berends--Giele current, and the
second line due to the non-linear part of the field-strength $\cf_P^{mn}$ reproduces
the quartic vertex \quarticV. This is natural since the quartic interaction in the
YM Lagrangian
arises from the non-linear part
of the field-strength.  Together with the single-particle case $\ce_i^m = J_i^m =
e_i^m$, the matching of \regroup\ at $\chi^\alpha_j = 0$ with the Berends--Giele
recursion \BGrecur\ completes the inductive proof of \repcurrent.

Also note that the recursion \rectwo\ for ${\cal X}_P^\alpha$ amounts to a
resummation of Feynman diagrams incorporating both the fermion propagator $k_m
\g^m_{\a\b}/k^2$ and the cubic coupling of two fermions with a boson, in accordance with
the Berends--Giele method \BerendsME\ applied to ten-dimensional SYM theory.

\subsec Supersymmetric Berends--Giele amplitude from the pure spinor formula 
\par\subseclab\threetwo

\noindent The relation \repcurrent\ between the ten-dimensional Berends--Giele
current $\ce^m_P$ in superspace and its purely gluonic
counterpart $J^m_P$ is now extended to their
corresponding tree-level amplitudes: the pure
spinor formula \AYM\ versus the Berends--Giele formula \BGformula.

To see the relation, note that \compon\
can be rewritten as
\eqnn\componZ
$$\eqalignno{
\langle M_X M_Y M_Z \rangle &= (\ce_{[X,Y]} \cdot \ce_Z)
+  \ce_X^m ({\cal X}_Y \g_m {\cal X}_Z) -  \ce_Y^m ({\cal X}_X \g_m {\cal X}_Z) &\componZ\cr
&{}+ \half\sum_{RS=Z}\big[(\ce_R\cdot \ce_X)(\ce_S\cdot \ce_Y)
- (\ce_R\cdot \ce_Y)(\ce_S\cdot \ce_X)\big]\ ,
}$$
provided that transversality \eoms\ and momentum conservation holds,
$k^m_X+ k^m_Y + k^m_Z=0$.
In particular, when $Z\rightarrow p+1$ is a single-particle label associated with
the $(p+1)^{\rm th}$ massless leg, the deconcatenation terms in the second line of \componZ\ vanish:
\eqn\componS{
\langle M_X M_Y M_{p+1} \rangle = (\ce_{[X,Y]} \cdot \ce_{p+1})
+  \ce_X^m ({\cal X}_Y \g_m {\cal X}_{p+1}) -  \ce_Y^m ({\cal X}_X \g_m {\cal X}_{p+1})\,.
}
Plugging the correlator \componS\ into the pure spinor superspace formula for
tree-level SYM amplitudes \AYM\ yields
\eqn\welldefined{
A^{\rm SYM}(1,2, \ldots p,p+1) =
\!\!\!\! \sum_{XY=12\ldots p}\!\Bigl[ (\ce_{[X,Y]} \cdot \ce_{p+1})
+  \ce_X^m ({\cal X}_Y \g_m {\cal X}_{p+1}) -  \ce_Y^m ({\cal X}_X \g_m {\cal X}_{p+1})
\Bigr] \,.
}
Alternatively, using \reczero\ and \eoms\ to identify $\ce^m_{12 \ldots p}$ and $\cX^\alpha_{12
\ldots p}$, this can be written as
\eqn\getAYM{
A^{\rm SYM}(1,2, \ldots p,p+1) =
s_{12 \ldots p} (\ce_{12 \ldots p}\cdot \ce_{p+1}) + k^m_{12 \ldots p} (\cX_{12
\ldots p}\g_m \cX_{p+1})\ .
}
In view of \repcurrent, the expression \getAYM\ reproduces the gluonic 
Berends--Giele formula \BerendsME\ in absence of fermions,
\eqn\moreill{
A^{\rm SYM}(1,2 \ldots, p,p+1) \big|_{\chi_j=0} = s_{12 \ldots p}(J_{12 \ldots p}
\cdot J_{p+1}) = A^{\rm YM}(1,2 \ldots, p,p+1)\,,
}
and additionally provides its supersymmetric completion.
Note that the bosonic currents $\ce^m_P$ contain even powers of gluino
wavefunctions $\chi_i^\alpha$ from the last term in \recone\ such as $s_{12} \ce_{12}^m=
s_{12} J^m_{12} + (\chi_1 \g^m \chi_2)$. Hence, both classes of terms on the right hand side of \getAYM\
contribute to fermionic amplitudes.

\subsec Divergent propagators and their cancellation
\par\subseclab\threefour

\subsubsec In components

\noindent From the definition \reczero\ it follows that both of $\ce^m_{P} $ and
${\cal X}^\alpha_P$ in \getAYM\ are proportional to a divergent propagator since
$s_{P}=0$ for a massless $(p+1)$-point amplitude. As well known from the
Berends--Giele formula for gluons \BerendsME, this is compensated by the formally
vanishing numerator containing $s_P=0$ in \BGformula. The same is true for its
supersymmetric completion derived in \getAYM\ since $k_P^m (\g_m
\cX_{p+1})_\alpha=0$ using $k^m_P = -k^m_{p+1}$ and the massless Dirac equation. The
interpretation is also the same; $s_{P}$ is the inverse of the bosonic propagator
$1/\partial^2$ while $k^P_m \g^m_{\a\b}$ is the inverse of the fermion propagator
$\partial_m \g^m_{\a\b}/\partial^{2}$.

\subsubsec In pure spinor superspace

The supersymmetric way to cancel a divergent propagator relies on the action of the
pure spinor BRST charge $Q\equiv \l^\alpha D_\alpha$ \psf\ on the currents $M_P$ \nptFT,
\eqn\deconcat{
E_P \equiv Q M_P = \sum_{XY=P}    M_X M_Y\,.
}
The integration of schematic form $\langle \l^3 \t^5 \rangle=1$ annihilates
BRST-exact expressions \psf. Because the single-particle superfield
$M_{p+1}$ is BRST closed, $QM_{p+1}=0$,
the superspace representation of
tree-level amplitudes in \AYM\ would be BRST exact $Q(M_P M_{p+1})$ if the current $M_{P}$ was
well defined in the phase space of $p+1$ massless particles \nptFT. However,
$M_P \sim
1/s_{P}$ and therefore the vanishing of $s_P$ prevents the amplitude
from being BRST exact. Just like \welldefined, the expression $\langle
\sum_{XY=P}M_XM_Y M_{p+1}\rangle$ does not contain any divergent propagator.

The assessment of BRST-exactness for a given superfield will play an important role in the derivation of
BCJ relations in section \fiveone.

\subsec Short representations and BRST integration by parts
\par\subseclab\threethree

\noindent
At first sight
the Berends--Giele formula \BGformula\ requires the $p$-current
$J^m_{12 \ldots p}$ in the computation of the $(p+1)$-gluon amplitude.
However, a diagrammatic method has been used by Berends and Giele in \BerendsHF\ to
obtain ``short'' representations of bosonic amplitudes up to eight points
which required no more than the
four-current and led to manifestly cyclic formul{\ae} for $A^{\rm YM}(1,2, \ldots,p+1)$. For
example, the six-point amplitude was found to be
\eqnn\shortAsix
$$\eqalignno{
A^{\rm YM}(1,2, \ldots,6) &= \half s_{123}J^m_{123}J^m_{456} +{1\over 3}[J_{12},J_{34}]^m J^m_{56}
&\shortAsix\cr
&{}+\half\{J_1,J_{23},J_{4}\}^mJ^m_{56}+ \{J_1,J_2,J_{34}\}^m J^m_{56}
+\cyclic{123456} \ , \cr
}$$
and similar expressions were written for the seven- and eight-point amplitudes
\BerendsHF.

In the framework of pure spinor superspace, the multiplicity of currents can be
shortened using integration by parts of the BRST charge. By \deconcat, this
amounts to
\eqn\cohom{
\sum_{XY=P}\langle M_X M_Y M_Q\rangle = \sum_{XY=Q}\langle M_P M_X M_Y\rangle\,,
}
which has been used in \nptFT\ to cast the superspace formula \AYM\ for $n$-point
trees into a manifestly cyclic form without any current of multiplicity higher than
${n \over 2}$, e.g.
\eqn\BRSTsix{
A^{\rm SYM}(1,2, \ldots,6)= {1\over3} \langle M_{12} M_{34} M_{56} \rangle
           +{1\over 2} \langle M_{123} (M_{45} M_{6} + M_4 M_{56}) \rangle +\cyclic{123456}  \ .
}
In terms of the components $\cm_{X,Y,Z}$ from the evaluation \compon\ of pure spinor superspace
expressions,
the component expressions for amplitudes of multiplicity $\leq 8$ are given by
\eqnn\shortMsix
$$\eqalignno{
A^{\rm SYM}(1,2, \ldots,4) &= {1\over 2} \cm_{12,3,4} + \cyclic{12 \ldots 4}  &\shortMsix\cr
A^{\rm SYM}(1,2, \ldots,5) &=  \cm_{12,3,45} + \cyclic{12 \ldots 5}  \cr
A^{\rm SYM}(1,2,\ldots ,6) &=   {1\over3}\cm_{12,34,56} 
           +{1\over 2} (\cm_{123,45,6} + \cm_{123,4,56} )    + \cyclic{12 \ldots 6} \cr
A^{\rm SYM}(1,2,\ldots ,7) &=  \cm_{123,45,67} + \cm_{1,234,567} + \cyclic{12\ldots 7}  \cr
A^{\rm SYM}(1,2,\ldots ,8) &= {1\over 2} (\cm_{1234,567,8} + \cm_{1234,56,78} +\cm_{1234,5,678}  )\cr
&\quad{}+ \cm_{123,456,78} +\cyclic{12\ldots 8} \ ,
}$$
see \nptFT\ for the nine- and ten-point analogues. Given the recursive nature of
the definitions of $\ce^m_P$, $\cf^{mn}_P$ and $\cX^\a_P$, the full component expansion of
the above amplitudes is readily available and reproduce the results available on
the website \WWW.

Note that the manipulations leading to \componS\ rely on a single-particle current
$M_{p+1}$ and therefore do not apply to the $\cm_{X,Y,Z}$ in \shortMsix.

\subsec The generating series of tree-level amplitudes

The way how component amplitudes \getAYM\ of SYM descend from the 
pure spinor superspace expression \AYM\ can be phrased
in the language of generating series. The solution
\eqn\sol{
\Bbb V \equiv \lambda^\alpha \Bbb A_\alpha =  \sum_i M_i t^i
+ \sum_{i,j} M_{ij} t^i t^j +\sum_{i,j,k} M_{ijk} t^i t^j t^k + \cdots
}
of the non-linear SYM equations \SYMeom\ generates color-dressed SYM amplitudes
via\foot{The representations of SYM amplitudes generated
by ${\rm Tr}\langle \Bbb V \Bbb V \Bbb V \rangle$ are related
to \AYM\ by BRST integration by parts \cohom.} \MafraGIA
\eqn\VVV{
{1\over 3}{\rm Tr}\langle \Bbb V \Bbb V \Bbb V \rangle =
\sum_{n=3}^{\infty} {n-2\over n}\!\!\!\! \sum_{i_1,i_2,\ldots,i_n}\!\!\!\! {\rm Tr}(t^{i_1} t^{i_2} \ldots t^{i_n}) A^{{\rm SYM}}(i_1,i_2,\ldots,i_n)  \ .
}
Note from \zerocomp\ that $\ce^m_P, {\cal X}_P^\alpha$ and $\cf^{mn}_P$
are just the $\theta=0$ components of the corresponding generating series
$\Bbb A^m$, $\Bbb W^\alpha$ and $\Bbb F^{mn}$.
Therefore \compon\ implies that
\eqnn\backtoFF
$$\eqalignno{
{1\over 3}{\rm Tr}\langle \Bbb V \Bbb V \Bbb V \rangle &=
{1\over 4}{\rm Tr} ( [\Bbb A_m , \Bbb A_n]\Bbb F^{mn})
+ {\rm Tr} (  \Bbb W \gamma^m \Bbb A_m  \Bbb W) \, \Big|_{\theta=0}  \cr
&= 
{\rm Tr} \Big( {1\over 4} \Bbb F_{mn} \Bbb F^{mn}
+ (\Bbb W  \gamma^m \nabla_m \Bbb W) \Big) \, \Big|_{\theta=0}\,. &\backtoFF
}$$
In passing to
the second line of \backtoFF, we have used the massless
Dirac equation $\nabla_m \gamma^m_{\alpha \beta} \Bbb W^{\beta}=0$ as
well as the field equation $\partial_m \Bbb F^{mn}=[\Bbb A_m,\Bbb F^{mn}]
+ \gamma^n_{\alpha \beta} \{ \Bbb W^\alpha , \Bbb W^\beta \}$ and discarded
a total derivative
to rewrite $(\partial_m \Bbb A_n) \Bbb F^{mn} =
- \Bbb A_n \big( [\Bbb A_m,\Bbb F^{mn}]
+ \gamma^n_{\alpha \beta} \{ \Bbb W^\alpha , \Bbb W^\beta \} \big)$.
The factor $1/3$ on the left-hand side of \backtoFF\ offsets the
sum over three terms that results from the cyclic symmetry of the
trace.

It is interesting to observe that the generating series of
tree-level amplitudes \backtoFF\ matches the ten-dimensional
SYM Lagrangian evaluated on the
generating series of (non-local) Berends--Giele currents
in superspace: $\Bbb F^{mn}(x,0)$ and $\Bbb W^\a(x,0)$.

\newsec BCJ relations from the cohomology of pure spinor superspace

In this section, we prove that the BCJ relations \BCJ\ among partial SYM amplitudes
follow from the vanishing of certain BRST-exact expressions in pure spinor
superspace and find a closed formula for them. A closely related property of tree
amplitudes is the possibility to express the complete kinematic dependence in terms
of $(n-2)!$ master numerators through a sequence of Jacobi-like relations \BCJ. A
superspace representation of such master numerators was given in \MafraKJ, and we
will provide a compact component evaluation along the lines of the previous
section.

\subsec Kleiss--Kuijf relations from symmetries of Berends--Giele currents
\par\subseclab\fivezero

\noindent For completeness,
we start by revisiting from a superspace perspective the Kleiss--Kuijf (KK) relations
among color-ordered amplitudes \KKref, firstly proven in \lanceKK.

The KK relations are conveniently described in the Berends--Giele framework. To see
this, recall that the superspace currents
$\cK_P\in \{\cA_\a^P, \cA^m_P,\cW^\a_P,\cF^{mn}_P\}$ satisfy
the symmetry property \EOMBBs
\eqn\alternal{
\cK_{A\shuffle B} = 0, \quad \forall \, A, B\neq\emptyset \ ,
}
see appendix B of \companion\ for a proof.
The symmetry \alternal\ of course also holds for
theta-independent components $\{
\ce^m_P, \cX^\a_P, \cf^{mn}_P\}$ of $\cK_P$, see \zerocomp.
Since the currents $\ce^m_P$ reduce to $J^m_P$ via \repcurrent, this is consistent with
the symmetry $J^m_{A\shuffle B} = 0$, $\forall A, B\neq\emptyset$ derived by
Berends and Giele in \BerendsZN.
The symmetry \alternal\ together with the identity\foot{Incidentally, the identity
\BGFromshuffle\ shows the equivalence between the statements given in equation (2)
of \BGschocker\ and Theorem 2.2 of \ree.}
\eqn\BGFromshuffle{
\cK_{B1A} - (-1)^{|B|}\cK_{1(A\shuffle B^T)} = - \sum_{XY=B}
(-1)^{|X|}\cK_{X^T \shuffle (Y 1 A)} 
-(-1)^{|B|}\cK_{B^T \shuffle (1 A)} 
\ ,
}
where $B^T$ denotes the reversal of the word $B$,
lead to an alternative form of \alternal,
\eqn\BGKK{
\cK_{B1A} - (-1)^{|B|}\cK_{1(A\shuffle B^T)} = 0 \ .
}
Since $E_P\equiv QM_P$ generalizes \BGKK\ to ${\cal K}_P \rightarrow E_P$,
the tree-level amplitude representation\foot{We omit the superscript
from $A^{\rm SYM}$ and write the labels as a subscript to avoid cluttering.}
\getAYM\ $A_{12 \ldots n}=\langle E_{12 \ldots n-1}M_n\rangle$
immediately yields the Kleiss--Kuijf relations
\eqn\KKrel{
A_{C1Bn} - (-1)^{|C|}A_{1(B\shuffle C^T)n} =
\langle  \big( E_{C1B} -
(-1)^{|C|}E_{1(B\shuffle C^T)}\big)M_n \rangle = 0\ ,
}
which reduce the number of independent color-ordered amplitudes
to $(n-2)!$ \KKref.

\subsec BCJ relations from the BRST cohomology
\par\subseclab\fiveone

\subsubsec Berends--Giele currents in BCJ gauge

\noindent There is a method to
construct Berends--Giele currents from quotients of local superfields $K_{[P,Q]}$ by
Mandelstam invariants whose precise form follows from an intuitive mapping with
cubic graphs (or planar binary trees) \refs{\EOMBBs,\companion}. For example,
the Berends--Giele currents associated with the local superfield
$V_{[P,Q]}\equiv \l^\a A_\a^{[P,Q]}$ up to multiplicity four are given by
\eqnn\MsfromVs
$$\displaylines{
M^{{\rm BCJ}}_{12} = {V_{[1,2]} \over s_{12}}, \quad \quad
M^{{\rm BCJ}}_{123} = {V_{[[1,2],3]} \over s_{12} s_{123}} + { V_{[1,[2,3]]} \over
s_{23} s_{123}}\,,\hfil\MsfromVs\hfilneg\cr
M^{{\rm BCJ}}_{1234} = {1 \over s_{1234}} \Big( { V_{[[[1,2],3],4]}\over s_{12}s_{123} }
 + {V_{[[1,[2,3]],4]}\over s_{23}s_{123} } 
 + { V_{[[1,2],[3,4]]}\over s_{12}s_{34} }
 + { V_{[1,[2,[3,4]]]} \over s_{34}s_{234} } 
 + { V_{[1,[[2,3],4]]} \over s_{23}s_{234} } \Big) \,.
}$$
As discussed in a companion paper \companion, one can perform a multiparticle
gauge transformation (denoted BCJ gauge) which enforces the superfields 
\eqn\Vbracket{
V_{123\ldots p} \equiv V_{[[ \ldots [[1,2],3], \ldots],p]}
}
in \MsfromVs\ with diagrammatic interpretation shown in \figMulti\ 
to satisfy the Lie symmetries of nested
commutators $[[ \ldots [[t^1,t^2],t^3],\ldots],t^p]$, e.g.
\eqn\Liesym{
V_{12} + V_{21} = 0,\quad
V_{123} + V_{231} + V_{312} = 0\,.
}

\ifig\figMulti{The tree diagram with an off-shell leg is represented
by the local superfield \Vbracket.}
{\epsfxsize=.60\hsize\epsfbox{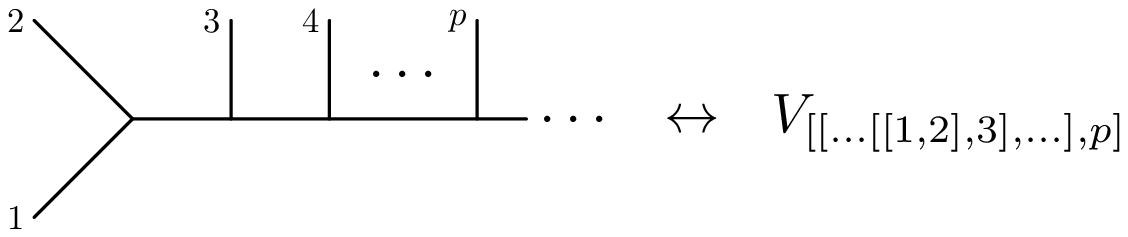}}

\noindent Moreover, BCJ gauge allows to reduce any other topology of bracketings to the master topology \Vbracket\ 
by a sequence of Jacobi-like identities 
\eqn\JacV{
V_{\ldots [[P,Q],R]\ldots}+V_{\ldots [[Q,R],P]\ldots}+V_{\ldots [[R,P],Q]\ldots}=0\,,
\quad \hbox{e.g. $V_{[[1,2],[3,4]]}=V_{1234}-V_{1243}$}\,.
}
Hence, the Berends--Giele current  $M^{\rm BCJ}_{12 \ldots p}$ can be expanded in terms of the $(p-1)!$
independent permutations of $V_{12 \ldots p}$. 
This is the same number of independent components as left by the Berends--Giele symmetry \alternal\ (here
for $\cK_{12 \ldots p}\rightarrow M^{\rm BCJ}_{12 \ldots p}$). As a crucial feature of Berends--Giele currents 
in BCJ gauge, there is an invertible mapping between the local superfields $V_{12\ldots p}$ and
$M^{\rm BCJ}_{12 \ldots p}$. More explicitly, for multiplicity
$p \leq 4$ one can use \Vbracket\ and \JacV\ to invert \MsfromVs\ and obtain
\eqnn\BGinv
$$\eqalignno{
V_{12}   &=  s_{12} M^{{\rm BCJ}}_{12} ,\qquad
V_{123} = s_{12}  (s_{23} M^{{\rm BCJ}}_{123}-s_{13} M^{{\rm BCJ}}_{213}) \,,\hfil  &\BGinv  \hfilneg\cr
V_{1234} &=  s_{12}\big[ s_{23} s_{34}M^{{\rm BCJ}}_{1234}  - s_{13} s_{34} M^{{\rm BCJ}}_{2134} + s_{14} s_{23} M^{{\rm BCJ}}_{3214} - s_{13} s_{24} M^{{\rm BCJ}}_{3124}\cr
&\qquad{} +s_{23}s_{24}(M^{{\rm BCJ}}_{1234}
+M^{{\rm BCJ}}_{1243}) - s_{13} s_{14}(M^{{\rm BCJ}}_{2134}
+M^{{\rm BCJ}}_{2143}) \big] 
\, \ .
}$$
The generalization to arbitrary rank can be read off from the
formula \nptTreeA
\eqn\trading{
{ V_{12\ldots p} \over z_{12} z_{23} \cdots z_{p-1,p}}  + {\rm perm}(2,\ldots,p)
= \prod_{k=2}^p
\sum_{m=1}^{k-1} { s_{mk} \over z_{mk} } M^{\rm BCJ}_{12\ldots p}  + {\rm perm}(2,\ldots,p)\,,
}
using partial fraction relations\foot{Note that
$Z_{12 \ldots p-1,p}\equiv 1/(z_{12}z_{23} \ldots z_{p-1,p})$
satisfies $Z_{A\shuffle B}=0, \forall A,B\neq\emptyset$.
} among the denominators made of $z_{ij} \equiv z_i - z_j$.

It is important to stress that the left-hand sides in \BGinv\ are {\it local}
expressions; all the kinematic poles in Mandelstam invariants cancel out from
the linear combinations of currents on the right-hand side. The poles cancel only when
the superfields are in the BCJ gauge. As we will see below,
this fact can be exploited to derive the BCJ relations \BCJ\ among color-ordered amplitudes.

\subsubsec Four- and five-point BCJ relations

\noindent
We shall now connect superfields in BCJ gauge with BCJ relations among partial SYM amplitudes. At the four-
and five-point level, one multiplies the local expressions in
\BGinv\ by a single-particle $V_n$ (which is BRST closed)
and uses the vanishing of BRST-exact expressions under the pure spinor bracket prescription $\langle \ldots \rangle$ \psf.
For example,
\eqn\BCJc{
{V_{123}\over s_{12}} = s_{23} M^{{\rm BCJ}}_{123} - s_{13} M^{{\rm BCJ}}_{213}  \ \ \Rightarrow \ \ 0 = \Big\langle 
Q \Big( {V_{123}\over s_{12}} V_4 \Big) \Big \rangle = \langle (s_{23} E^{{\rm BCJ}}_{123} - s_{13} E^{{\rm BCJ}}_{213}) V_4 \rangle}
with $ E^{{\rm BCJ}}_{P} \equiv QM^{{\rm BCJ}}_{P}$ corresponds to the
four-point\foot{The three-point BCJ relation $0=s_{12} A^{\rm SYM}(1,2,3)$
following from $s_{12}=0$ can be formally derived via $0=\langle Q V_{12} V_3
\rangle=s_{12} \langle V_1 V_2 V_3\rangle$.} BCJ relation \BCJ\ by \AYM,
\eqn\BCJd{
0=s_{23} A^{\rm SYM}(1,2,3,4) - s_{13} A^{\rm SYM}(2,1,3,4)  \ . 
}
Note that the BCJ gauge for the local superfields is a crucial requirement in this
derivation --- in a generic gauge, $s_{23} M_{123} - s_{13} M_{213}$ would be an
ill-defined expression containing divergent propagators of the form $1/{}s_{123}$
and the BRST triviality of $(s_{23} E_{123} - s_{13} E_{213})V_4$ would no longer
be guaranteed.

Similarly, the identities
\eqnn\QBCJa
$$\eqalignno{
{V_{1234}\over s_{12} s_{123}}+{V_{3214}\over s_{23} s_{123}} &=
s_{34} M^{{\rm BCJ}}_{1234}  + s_{14} M^{{\rm BCJ}}_{3214} 
- s_{24} (M^{{\rm BCJ}}_{1324} + M^{{\rm BCJ}}_{3124}) &\QBCJa \cr
{V_{1234} - V_{1243} \over s_{12}s_{34}} &= s_{23} M^{{\rm BCJ}}_{1234}
- s_{13} M^{{\rm BCJ}}_{2134}  - s_{24} M^{{\rm BCJ}}_{1243}
+ s_{14} M^{{\rm BCJ}}_{2143}
}
$$ 
derived from \BGinv\ with manifestly well-defined left-hand side imply the BCJ relations \BCJ
\eqnn\QBCJe
$$\eqalignno{
0 = \Big\langle 
Q \Big( {V_{1234}\over s_{12} s_{123}}+{V_{3214}\over s_{23} s_{123}} \Big) V_5 \Big \rangle&=
s_{34} A^{\rm SYM}(1,2,3,4,5)  + s_{14} A^{\rm SYM}(3,2,1,4,5) &\QBCJe\cr
& \ \ \ - s_{24} \big[A^{\rm SYM}(1,3,2,4,5) + A^{\rm SYM}(3,1,2,4,5)\big]  \cr
0 = \Big\langle 
Q \Big(
{V_{1234} - V_{1243} \over s_{12}s_{34}} \Big) V_5 \Big \rangle&=
s_{23} A^{\rm SYM}(1,2,3,4,5) - s_{13} A^{\rm SYM}(2,1,3,4,5)  \cr
& \ \ \  - s_{24} A^{\rm SYM}(1,2,4,3,5)  + s_{14} A^{\rm SYM}(2,1,4,3,5) \ .  \cr
}
$$
Even though the above derivation relies on the choice of BCJ gauge, the
subamplitudes in the resulting BCJ relations are independent on the
multiparticle gauge for the currents $M_P$. This can be seen from
the non-linear gauge invariance in the generating series \VVV\ of
the amplitude formula \AYM.

\subsubsec Higher-point BCJ relations

Along the same lines, one can verify in a basis of $V_P$ that the expression \EOMBBs
\eqn\QEone{
M^{{\rm BCJ}}_{S[A,B]} \equiv \sum_{i=1}^{|A|} \sum_{j=1}^{|B|} (-1)^{i-j+|A|-1}
s_{a_i b_j} M^{{\rm BCJ}}_{(a_1 a_2\ldots a_{i-1} \shuffle a_{|A|} a_{|A|-1}\ldots a_{i+1})a_ib_j
(b_{j-1}\ldots b_2 b_1 \shuffle b_{j+1} \ldots b_{|B|})} 
}
with $A=a_1 a_2\ldots a_{|A|}$ and $B=b_1 b_2\ldots b_{|B|}$ does not have any pole
in $s_{AB}$. One can therefore identify the following BRST-exact combinations of
$(|A|+|B|+1)$-point amplitudes,
\eqnn\QBCJf
$$\eqalignno{
0 &= (-1)^{|A|-1} \langle Q(M^{{\rm BCJ}}_{S[A,B]} M_n) \rangle &\QBCJf \cr
&= \sum_{i=1}^{|A|} \sum_{j=1}^{|B|} (-1)^{i-j} s_{a_i b_j} \langle 
E^{{\rm BCJ}}_{(a_1 a_2  \ldots a_{i-1} \shuffle a_{|A|} a_{|A|-1}\ldots a_{i+1})a_ib_j
(b_{j-1}\ldots b_2 b_1 \shuffle b_{j+1} \ldots b_{|B|})} M_n \rangle \cr
&= \sum_{i=1}^{|A|} \sum_{j=1}^{|B|} (-1)^{i-j} s_{a_i b_j}
A^{\rm SYM}(
(a_1 \ldots a_{i-1} \shuffle a_{|A|} \ldots a_{i+1}),a_i,b_j,
(b_{j-1}\ldots   b_1 \shuffle b_{j+1} \ldots b_{|B|})
,n) \ ,
}
$$
which all boil down to BCJ relations in some representation \refs{\BCJ,
\BjerrumBohrRD, \StiebergerHQ, \ChenJXA}. For the single-particle
choice $A=1$ along with $B=2,3,4,\ldots, (n-1)$, \QBCJf\ reduces
to the fundamental BCJ relations
\eqnn\QBCJg
$$\eqalignno{
 0 &= - \langle Q(M^{{\rm BCJ}}_{S[1,234\ldots (n-1)]} M_n) \rangle &\QBCJg\cr
 &= s_{12} A^{\rm SYM}(2,1,3,\ldots,n) + (s_{12}+s_{13})  A^{\rm SYM}(2,3,1,4,\ldots,n) \cr
 & \ \ \ + \cdots+ (s_{12}+s_{13}+\ldots+s_{1,n-1}) A^{\rm SYM}(2,3,\ldots,n-1,1,n) \ ,
}
$$
which are well-known to leave $(n-3)!$ independent
subamplitudes \refs{\BCJ, \BjerrumBohrRD, \StiebergerHQ, \ChenJXA}.

\subsec Component form of BCJ numerators
\par\subseclab\fiveseven

\noindent The initial derivation of BCJ relations in \BCJ\ relied on the duality
between color and kinematics, i.e. the existence of particular representations of
tree amplitudes. The functions of polarizations and momenta associated with the
cubic graphs in such a ``BCJ representation'' are assumed to obey the same Jacobi
identities as the color factors made of structure constants $f^{abc}$ of the gauge
group. As a consequence, the complete information on polarizations and momenta
reside in $(n-2)!$ master graphs which can be chosen to be the half-ladder diagrams
with fixed endpoints $1$ and $n-1$ as depicted in \figmaster\ and arbitrary
permutations of the remaining legs $2,3,\ldots,n-2$ and $n$.

\ifig\figmaster{The $(n-2)!$ half-ladder diagrams with legs $1$ and $n-1$ attached
to opposite endpoints encode the complete kinematic dependence
in a BCJ representation.}
{\epsfxsize=.60\hsize\epsfbox{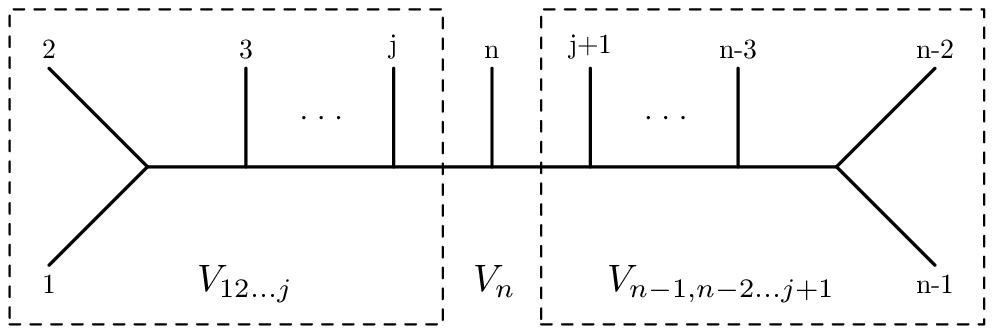}}

An explicit realization of the BCJ duality for tree-level amplitudes was given in
\MafraKJ\ based on the tree amplitudes of the pure spinor superstring. The master
graphs in the figure were associated with local kinematic numerators\foot{Note that
the precursors of $V_{12\ldots p}$ were denoted by $T_{12\ldots p}$ in \MafraKJ.}
$\langle V_{12\ldots j} V_{n-1,n-2\ldots j+1} V_n \rangle$ labeled by
$j=1,2,\ldots,n-2$ along with the $(n-3)!$ permutations of the legs
$2,3,\ldots,n-2$. The kinematic factors for any other graph can be reached by a
sequence of Jacobi relations, and this representation agrees with the field-theory
limit of the open superstring amplitude, i.e. yields the right SYM amplitude.

The techniques of \companion\ (in particular the discussion of BCJ/HS gauge)
give rise to a compact formula for their components,
\eqn\BCJcom{
\langle V_{A} V_{B} V_C \rangle = 
 \half e^m_A f^{mn}_B e^n_C  + (\chi_A \g_m \chi_B) e_C^m + \cyclic{ABC} \ ,
}
whose form is completely analogous to \compon. The constituents $e^m_A, f^{mn}_A$ and
$\chi^\alpha_A$ of \BCJcom\ are {\it local\/} multiparticle polarizations
and will be explained below.

\subsubsec Local multiparticle polarizations

The discussion of recursion relations for local superfields given in \companion\
has a direct counterpart for their multiparticle
polarizations $e^m_A, f^{mn}_A$ and $\chi^\alpha_A$ which constitute
their theta-independent terms.
The setup
starts with a recursive definition for local multiparticle
polarizations $\hat e^m_A, \hat f^{mn}_A$
and $\hat \chi^\alpha_A$ whose labels do not satisfy the
symmetries of a Lie algebra, for example $\hat e^m_{123} + \hat e^m_{231}
+ \hat e^m_{312} \neq 0$ (their hatted notation is a reminder of this symmetry
failure). However, non-linear gauge variations of their multiparticle superfields can be exploited to find 
a gauge where the symmetries are indeed satisfied.

The recursive definition of the hatted components is given by
\eqnn\localBG
$$\eqalignno{
\hat e^{m}_{12\ldots p}  &= - {1 \over 2 }  \bigl[ \hat e_{12\ldots p-1}^m (k_{12\ldots p-1}\cdot  \hat e_p) +\hat e_{12\ldots p-1}^n   \hat f_p^{mn}
- ( \hat \chi_{12\ldots p-1}\g^m \hat \chi_p)
- (12\ldots p-1 \leftrightarrow p)\bigr] \cr
\hat \chi^{\alpha}_{12\ldots p} &= 
{1 \over 2 } k_{12\ldots p}^n \g_n^{\alpha \beta} 
\big[ \hat e_{12\ldots p-1}^m (\g_m\hat \chi_p)_\beta  - (12\ldots p-1 \leftrightarrow p) \big] \ , &\localBG
}$$
and it starts with $\hat e_i^m = e_i^m$ and $\hat \chi^\alpha_i = \chi^\alpha_i$.
The local field-strength is defined~by
\eqn\fstr{
\hat f^{12 \ldots p}_{mn} \equiv k^{12 \ldots p}_m \hat e^{12 \ldots p}_n - k^{12 \ldots p}_n \hat e^{12 \ldots p}_m
+ \sum_{j=2}^{p} \sum_{\d \in P(\beta_j)}  (k_{12 \ldots j-1}\cdot k_j)\,
\hat e_{[n}^{12{\ldots} j-1,\{\d\}}\,  \hat e_{m]}^{j, \{\beta_j \backslash \d \} }\,,
}
with shorthand $\beta_j = \{j+1,j+2,\ldots,p\}$ and $P(\beta_j)$ denoting the power set of $\beta_j$, e.g.
\eqnn\fieldex
$$\eqalignno{
 \hat f^{mn}_{1} &=  f^{mn}_{1} = k^m_1 e^n_1 - k^n_1 e^m_1  \ , \ \ \ \ \hat f^{mn}_{12} =f^{mn}_{12} = k^m_{12}  e^n_{12} - k^n_{12}  e^m_{12} - s_{12} e_1^{[m} e_2^{n]}  \cr
\hat f^{mn}_{123} &=  k^m_{123}  \hat e^n_{123} - k^n_{123} \hat e^m_{123}
- (s_{13}+s_{23}) e^{[m}_{12} e^{n]}_3
 - s_{12} ( e^{[m}_{1} e^{n]}_{23} - e^{[m}_{2} e^{n]}_{13})   \ .  &\fieldex
}$$
Up to and including multiplicity $p=2$, the multiparticle polarizations in the BCJ
numerators \BCJcom\ agree with their hatted counterparts in \localBG,
\eqnn\BCJtwo
$$\eqalignno{
e^{m}_{12} &= \hat e^{m}_{12} = e^m_2 (e_1\cdot k_2) - e^m_1 (e_2\cdot k_1) +
\half(k_1^m-k_2^m)(e_1\cdot e_2) + (\chi_1 \gamma^m \chi_2) &\BCJtwo\cr
 \chi^{\alpha}_{12} &=  \hat \chi^\alpha_{12} = {1\over 2} k_{12}^p \gamma_p^{\alpha \beta} \big[ e_1^m (\gamma_m \chi_2)_\beta - e_2^m (\gamma_m \chi_1)_\beta \big]
 \ ,
}$$
while multiplicities $p \geq 3$ require redefinitions $\hat h_{12\ldots p}$ starting with
\eqn\BCJthree{
e^{m}_{123} = \hat e^{m}_{123} - k_{123}^m \hat h_{123} \ ,
\quad  \chi^{\alpha}_{123} =  \hat \chi^\alpha_{123} \ .
}
The redefinition of $\hat e^m_{123}$ in \BCJthree\ ensures the Lie symmetry
$e^{m}_{123}+e^{m}_{231}+e^{m}_{312}=0$.
At multiplicity $p=4$, we have
\eqnn\BCJfour
$$\eqalignno{
e^{m}_{1234} &= \hat e^{m}_{1234} + (k_{123} \! \cdot \! k_4) \hat h_{123} e_4^m - (k_{12} \!\cdot \!k_3) \hat h_{124} e_3^m  - (k_1\! \cdot \!k_2) (\hat h_{134} e_2^m - \hat h_{234} e_1^m) - k_{1234}^m \hat h_{1234} 
\cr
\chi^{\alpha}_{1234} &=  \hat \chi^\alpha_{1234} + (k_{123} \!\cdot \! k_4) \hat h_{123} \chi_4^\alpha - (k_{12} \! \cdot \! k_3) \hat h_{124} \chi_3^\alpha - (k_1 \! \cdot \! k_2) (\hat h_{134} \chi_2^\alpha - \hat h_{234} \chi_1^\alpha) \ , &\BCJfour
}$$
and the rank-five example can be extracted from \companion\ as will be 
explained shortly. The scalar correction terms $\hat h_{12\ldots p}$
in \BCJthree\ and \BCJfour\ can be reduced to building blocks 
\eqn\habc{
h_{A,B,C}\equiv {1\over 4} e_A^m  f_B^{mn} e_C^n +{1\over 2}(\chi_A \gamma_m \chi_B) e^m_C + {\rm cyclic}(ABC)
}
made of multiparticle polarizations of lower multiplicity $\leq p-2$ via
\eqnn\hversushat
$$\eqalignno{
3\hat h_{123} &\equiv   h_{1,2,3} 
\cr
4\hat h_{1234} &\equiv h_{12,3,4} + h_{34,1,2} - {1\over 2} h_{1,2,3} (k_{123}\cdot e_4) &\hversushat \cr
&  + {1\over 6}\big[ h_{1,3,4} (k_{134}\cdot e_2) - h_{2,3,4} (k_{234}\cdot e_1) - h_{1,2,4} (k_{124} \cdot e_3) \big]
\ .
}$$
Once the redefinition $e^m_{12\ldots p} = \hat e^m_{12\ldots p} + \ldots$ for
the multiparticle polarization
has been performed, the corresponding ``unhatted'' field-strength relevant for
the BCJ numerators
in \BCJcom\ is obtained completely analogously to \fstr,
\eqn\fstrnohat{
f^{12 \ldots p}_{mn} \equiv k^{12 \ldots p}_m e^{12 \ldots p}_n
- k^{12 \ldots p}_n  e^{12 \ldots p}_m
+ \sum_{j=2}^{p} \sum_{\d \in P(\beta_j)}  (k_{12 \ldots j-1}\cdot k_j)\,
e_{[n}^{12{\ldots} j-1,\{\d\}}\,   e_{m]}^{j, \{\beta_j \backslash \d \} } \ .
}

\subsubsec Higher multiplicity

As already mentioned, the above
redefinitions of $\hat e^m_{12\ldots p}$, $\hat \chi^\alpha_{12\ldots p}$ and
$\hat f^{mn}_{12\ldots p}$ descend from the superspace discussion in section~3
of \companion.
In particular, the corrections $h_{A,B,C}$ defined
in \habc\ are the $\theta=0$ component of a local superfield $H_{A,B,C}(x,\t)$ which
was completely specified up to multiplicity five in \companion. So
the full expressions of
$e^m_{12345},\chi^\alpha_{12345}$ and $f^{mn}_{12345}$ are readily
available.

At the same time, there is no obstruction
to pushing these recursive constructions even further, leading to local
multiparticle polarizations
$e^m_{P},\chi^\alpha_{P}$ and $f^{mn}_{P}$ of higher multiplicity.
Therefore, together with the central formula \BCJcom\ for local components,
the discussion in this section provides access to the supersymmetric
components of the {\it local} BCJ-satisfying numerators of \MafraKJ\
in a recursive fashion.

\newsec{Conclusion and outlook}

In this work, we have extracted and streamlined component information from
tree-level scattering amplitudes in pure spinor superspace. The results are based
on simplified theta-expansions for multiparticle superfields of ten-dimensional
SYM which are attained via non-linear gauge transformations in a companion paper
\companion.  More specifically:
\medskip
\item{$\bullet$} The $n$-point tree-level amplitude derived in \nptFT\ from
locality, supersymmetry and gauge invariance is shown to reproduce
the Berends--Giele formula, and the supersymmetrization by fermionic
component amplitudes is worked out.
\item{$\bullet$} BCJ relations are derived from the decoupling of
BRST-exact expressions in pure spinor superspace.
\item{$\bullet$} Kinematic tree-level numerators \MafraKJ\ satisfying the
BCJ duality between color and kinematics are translated into components.

\noindent The resulting ten-dimensional component amplitudes together with their
BCJ representations and dimensional reductions will have a broad range of
applications. With appropriate truncations of the gluon and gluino components, they
are suitable to determine $D$-dimensional unitarity cuts in a variety of theories including QCD, 
see e.g.~\refs{\EllisIR ,\DaviesVT, \BadgerGXA} and references therein.

It would be interesting to relate the multiparticle polarizations in the component
form of the BCJ numerators to the approach of \TolottiCAA. In that reference,
formally vanishing non-local terms are added to the Yang--Mills Lagrangian to 
automatically produce BCJ numerators. The interplay between Lagrangians and
generating series of kinematic factors might shed further light on the superfield
redefinitions in \companion\ underlying our BCJ numerators.

\bigskip \noindent{\bf Acknowledgements:}
We would like to express our gratitude to Seungjin Lee for collaboration
on the closely related paper \companion\ and for numerous insightful discussions.
We are also indebted to John-Joseph Carrasco for valuable comments on a draft of this article.
CRM wishes to acknowledge support from
NSF grant number PHY 1314311 and the Paul Dirac Fund. We acknowledge support by the
European Research Council Advanced Grant No. 247252 of Michael Green.
OS is grateful to DAMTP in Cambridge for kind hospitality during various
stages of this project, and OS additionally thanks the IAS in Princeton for kind
hospitality during completion of this work. CRM is grateful to AEI in Potsdam for
the warm hospitality during intermediate stages of this work and for partial
financial support.

\listrefs

\bye